\documentclass[twocolumn]{aastex62}
\usepackage{xspace}
\usepackage{graphicx}
\usepackage{footnote}
\usepackage{color}
\usepackage{hyperref}

\def\ltsima{$\; \buildrel < \over \sim \;$}
\def\simlt{\lower.5ex\hbox{\ltsima}}
\def\gtsima{$\; \buildrel > \over \sim \;$}
\def\simgt{\lower.5ex\hbox{\gtsima}}

\newcommand {\um}{$\mu$m}
\newcommand{\kpc}{{\rm\,kpc}}

\newcommand{\msun}{{\rm\,M$_\odot$}}

\newcommand{\lsun}{{\rm\,L$_\odot$}}

\newcommand{\mstar}{{\rm\,M$_\star$}}

\newcommand{\lir}{L$_{\rm IR}$}

\newcommand{\tdust}{T$_{\rm d}$}

\newcommand{\Mdust}{{M$_{\rm dust}$}}
\newcommand{\re}{R$_{\rm e}$}

\newcommand{\Dms}{{D$_{\rm MS}$}}
\newcommand{\e}[1]{$\times\,10^{#1}$}
\newcommand{\lpeak}{\rm\,$\lambda_{\rm peak}$}
\newcommand{\sfrsurfdens}{{\rm \lir\,\re$^{-2}$}}

\newcommand{\umin}{{$U_{\rm min}$}}
\newcommand{\umax}{{$U_{\rm max}$}}
\newcommand{\deltalogssfr}{$\Delta$log(SSFR)$_{\rm MS}$}

\shorttitle{Dust Temperatures in DSFGs}
\shortauthors{A. D. Burnham et al.}

\begin{document}

\title{The Physical Drivers of the Luminosity-Weighted Dust Temperatures in High-Redshift Galaxies}

\author{Anne D. Burnham}
\affil{Department of Astronomy, The University of Texas at Austin, 2515 Speedway Blvd Stop C1400, Austin, TX 78712, USA}
\affil{Department of Astronomy, Yale University, 52 Hillhouse Avenue, New Haven, CT 06511, USA}

\author{Caitlin M. Casey}
\affiliation{Department of Astronomy, The University of Texas at Austin, 2515 Speedway Blvd Stop C1400, Austin, TX 78712, USA}

\author{Jorge A. Zavala}
\affiliation{Department of Astronomy, The University of Texas at Austin, 2515 Speedway Blvd Stop C1400, Austin, TX 78712, USA}

\author{Sinclaire M. Manning}
\affiliation{Department of Astronomy, The University of Texas at Austin, 2515 Speedway Blvd Stop C1400, Austin, TX 78712, USA}

\author{Justin S. Spilker}
\altaffiliation{Hubble Fellow}
\affiliation{Department of Astronomy, The University of Texas at Austin, 2515 Speedway Blvd Stop C1400, Austin, TX 78712, USA}

\author{Scott C. Chapman}
\affiliation{Department of Physics and Atmospheric Science, Dalhousie University, 6310 Coburg Road, Halifax, Nova Scotia,
B3H 4R2, Canada\\
}

\author{Chian-Chou Chen}
\affiliation{Academia Sinica Institute of Astronomy and Astrophysics (ASIAA), No. 1, Section 4, Roosevelt Rd., Taipei 10617, Taiwan}

\author{Asantha Cooray}
\affiliation{Center for Cosmology, Department of Physics and Astronomy, University of California, Irvine, CA 92697, USA}

\author{David B. Sanders}
\affiliation{Institute for Astronomy, University of Hawai’i at Manoa, 2680 Woodlawn Drive, Honolulu, HI 96822, USA}

\author{Nick Z. Scoville}
\affiliation{California Institute of Technology, 1216 East California Boulevard, Pasadena, CA 91125, USA}

\begin{abstract}
The underlying distribution of galaxies' dust SEDs (i.e., their spectra re-radiated by dust from rest-frame $\sim$\,3\,\um\,--\,3\,mm) remains relatively unconstrained due to a dearth of FIR/(sub)mm data for large samples of galaxies. It has been claimed in the literature that a galaxy's dust temperature---observed as the wavelength where the dust SED peaks (\lpeak)---is traced most closely by its specific star-formation rate (sSFR) or parameterized ‘distance’ to the SFR--\mstar \ relation (the galaxy `main sequence'). We present 0\farcs24\ resolved 870\,\um \ ALMA dust continuum observations of seven $z=1.4-4.6$ dusty star-forming galaxies (DSFGs) chosen to have a large range of well-constrained luminosity-weighted dust temperatures. We also draw on similar resolution dust continuum maps from a sample of ALESS submillimeter galaxies from \citet{hodge16a}. We constrain the physical scales over which the dust radiates and compare those measurements to characteristics of the integrated SED. We confirm significant correlations of \lpeak\ with both \lir \ (or SFR) and $\Sigma_{\rm IR}$\ ($\propto$SFR surface density). We investigate the correlation between $\log_{10}$(\lpeak) and $\log_{10}$($\Sigma_{\rm IR}$) and find  the relation to hold as would be expected from the  Stefan-Boltzmann Law, or the effective size of an equivalent blackbody. The correlations of \lpeak \ with sSFR and distance from the SFR--\mstar\ relation are less significant than those for  $\Sigma_{\rm IR}$ or \lir; therefore, we conclude that the more fundamental tracer of galaxies' luminosity-weighted integrated dust temperatures are indeed their star-formation surface densities  in line with local Universe results, which relate closely to the underlying geometry of dust in the ISM.
\end{abstract}


\section{Introduction} \label{sec:intro}

Dusty Star-Forming Galaxies (DSFGs) have incredibly high star-formation rates and in the first few Gyr produce $\sim$\,$50\%$ of the stellar mass in the Universe \citep{casey12b, Gruppioni13a}. They are vitally important to galaxy evolution, but many of their fundamental dust characteristics are not well studied. Though their far-infrared through millimeter spectral energy distributions (SEDs) are relatively straightforward to interpret as a linear combination of modified blackbodies from dust, which re-radiates nascent starlight, the lack of detailed photometry along that SED (rarely exceeding a few measurements across a 1000\,\um \ range) has limited our understanding of the physics governing that dust in the interstellar medium (ISM). This includes the physical scales over which the dust radiates and how clumpy it is, perhaps tracing back to its origins in/around compact star clusters. Physically tracing the relationship between integrated SEDs and underlying geometry is critically important.

A major challenge in characterizing galaxies' dusty SEDs has been the fundamental limitation of infrared through radio datasets, particularly for high-$z$ galaxies. For ten years before 2010 most analysis was limited to SCUBA flux densities along the Rayleigh-Jeans (RJ) tail of blackbody emission, which made it impossible to constrain both the IR luminosity and dust temperature of a given high-$z$ galaxy. Even the vast improvement ushered in by \textit{Herschel} in the last decade has fundamental limitations given the telescope's large beamsize at 250\,\um$-$500\,\um, where SEDs of $z=1-3$ galaxies peak. The uncertainty brought on by confusion noise, added with the relatively shallow depth of \textit{Herschel} surveys provides some moderate breakthroughs in measuring galaxies' dust temperatures across cosmic time \citep[e.g.][]{symeonidis13a,lee13a}. 

Dust temperature is observationally constrained through measurement of \lpeak, the wavelength at which the SED peaks in the rest-frame, which is inversely proportional to the underlying physical dust temperature via Wien's law \citep{wien1897}. The precise mapping of \lpeak \ to \tdust \ depends on the underlying opacity of the dust in the ISM \citep*[see Figure 20 of][]{casey14a} and thus \tdust\ is usually unconstrained without spatially-resolved observations. Empirical datasets have shown us that DSFGs at all epochs have higher temperatures at higher IR luminosities \citep[e.g. Ulza, Perault 1987, \citealt*{blain04a};][]{casey14a}. This relationship is akin to a Stefan-Boltzmann law for the cold ISM on galaxy scales  even without direct accounting for sources' emitting regions or underlying dust opacity. Indeed, previous studies in the local universe see a relationship between the star-formation rate surface density, $\Sigma_{\rm IR}$, and integrated galaxy SED dust temperatures \citep{chanial07,lutz16a} as would be expected given a Stefan-Boltzmann type relation. This physical reasoning has also been used to interpret the integrated SEDs of high-redshift dusty galaxies \citep[e.g.][]{hodge16a,simpson17}  though some discrepancies in interpretation remain.

Trends in galaxies' dust temperatures at high redshifts have been measured frequently in the literature with somewhat mixed, potentially contradictory results. Some suggest evolution toward colder dust temperatures for higher redshift systems \citep{casey12b, symeonidis13a, lee13a, kirkpatrick17a} while others suggest evolution toward warmer dust temperatures at higher redshifts \citep[e.g.][]{magdis12a, magnelli14a, bethermin15a, bethermin17, schreiber18}. All of these works attempt to quantify the underlying physical drivers of galaxies' bulk dust temperatures. We describe the basis of these claims and how they are not actually contradictory to one another in a broader context in Section \ref{sec:context} of this paper. We use this diverse range of claims, and the physics used to explain them, as motivation for this study.

\begin{figure*}
\includegraphics[scale=0.62]{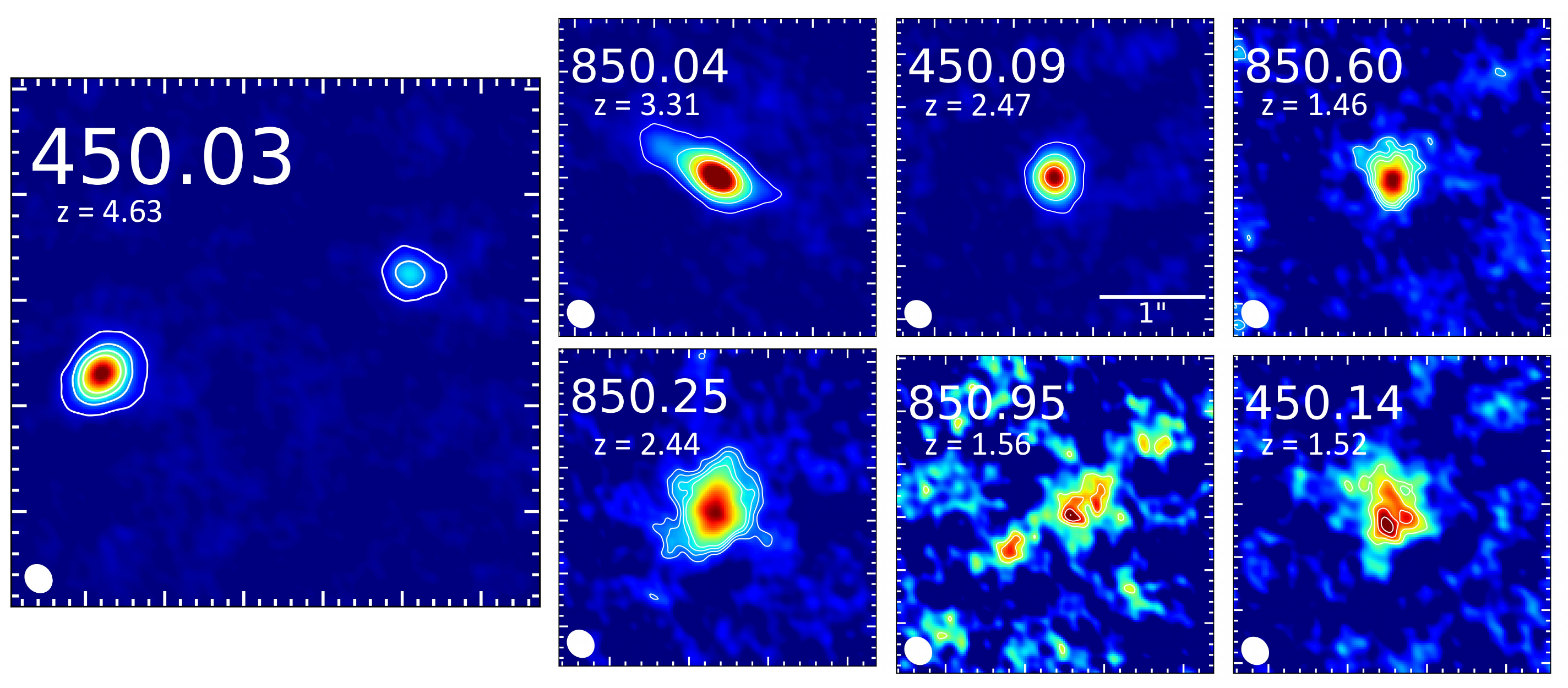}
\label{fig:dustmass}
\caption{ALMA 870\,\um \ images, proportional to the dust mass distribution for all sources except for perhaps 450.03, which sits at a much higher $z$. All images are $3\arcsec\times3\arcsec$, except for 450.03, which is $5\arcsec\times5\arcsec$ to show its two components \citep{jimenez20}. The ALMA beam is shown in the bottom left corner. For sources 450.03, 850.04, and 450.09, the contours start at $5\sigma$ and increment by $10\sigma$. For the remaining sources, the contours start at $3\sigma$ and increase by $1\sigma$.}
\end{figure*}

\begin{figure*}
\centering
\includegraphics[width=0.99\textwidth]{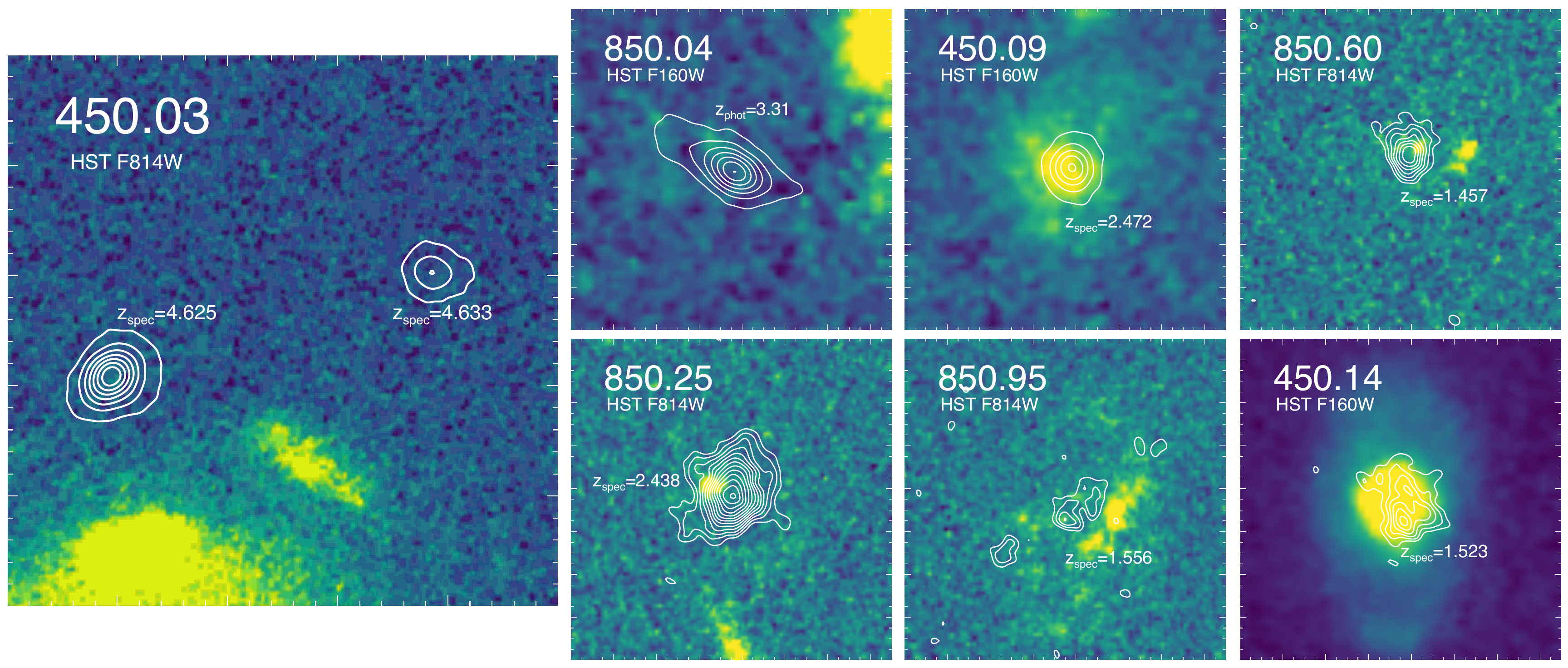}
\label{fig:oircutouts}
\caption{ Optical or Near-infrared imaging from {\it Hubble Space Telescope} of the COSMOS sample, with ALMA contour levels and same scale as in Figure~\ref{fig:dustmass}.  Where near-infrared imaging exists (from the CANDELS survey) it is shown (in the WFC3/F160W filter), otherwise $i$-band imaging from ACS/F814W is shown.  All sources here have secure spectroscopic redshifts (from either the millimeter or near-infrared) save 850.04, whose photometric redshift of $z=3.31$ is based on an OIR counterpart separated only by 0$\farcs$4 from the measured centroid of dust emission.  The ALESS sample has presented similar cutouts against {\it HST} imaging in \citet{chen15a}.}
\end{figure*}

In this paper, we have selected a sample of seven DSFGs with a wide range of measured dust temperatures and confirmed spectroscopic redshifts for high spatial-resolution submillimeter (submm) follow-up to directly test for correlation between dust temperature and other galaxy characteristic quantities. We use new resolved dust continuum maps from ALMA band 7 observations at 870\,\um \ of these seven DSFGs to investigate the hypothesis that dust temperature maps directly to other observable characteristics such as galaxy size, morphology, star-formation surface density, and sSFR. We also include eleven ALESS sources in our analysis from \citet{hodge16a}, whose observations and sample are similar, though their sources were 1.3$\times$ brighter and observations were slightly shallower. Section \ref{sec:context} provides background from the literature giving more in-depth motivation for this investigation. Section \ref{sec:sample} describes sample selection and observations. Section  \ref{sec:results} gives our calculations of key physical properties of the relevant DSFGs. In Section \ref{sec:discussion}, we discuss our findings, and we provide our conclusions in Section \ref{sec:conclusion}. We assume a concordance cosmology throughout this paper, adopting 
$H_{0}=70\,$km\,s$^{-1}$\,Mpc, $\Omega_{M}=0.3$, and $\Omega_{\Lambda}=0.7$.
Where SFRs are used, we assume a Kroupa IMF \citep{kroupa03a}  and scaling relations drawn from \citet{kennicutt12}.


\section{Context of Galaxies' Dust Temperatures}
\label{sec:context}

We draw motivation for this detailed study of a subsample of galaxies with high-quality resolved millimeter dust maps from the many discussions presented in the literature of trends in galaxies' overall dust temperatures with redshift. This section provides context of these discussions.

Dust is heated by a radiation field with an intensity represented by $U$. The dust is heated in two environments: the ambient radiation field heating the diffuse ISM and  discrete photon heating within star-forming regions. The diffuse ISM portion is heated by a radiation field with constant \umin, and the other portion is heated by radiation (primary and secondary photons) from young stars with intensities ranging from \umin\ and \umax\ \citep{magdis12a}. The equation governing the dust-weighted mean starlight intensity scale factor, $\langle\!U\!\rangle$ is defined in \citet{magdis12a}. From this equation, $\langle\!U\!\rangle$ is proportional to \lir/\Mdust. A corollary is that the luminosity-weighted dust temperature correlates directly with $\langle\!U\!\rangle$. The stronger the radiation field, the higher the dust temperature. Similarly, the more compact the dust around the source of incident radiation, the hotter the dust. 

\citet{magdis12a} used samples of \textit{Herschel}-detected galaxies to argue that the galaxy IR-spectral energy distribution depends solely on $\langle\!U\!\rangle$ and is independent of sSFR and \mstar. They also argue that $\langle\!U\!\rangle$ evolves with time such that main sequence galaxies at earlier epochs had more intense radiation fields, or higher $\langle\!U\!\rangle$. They find that \tdust \ evolves with time such that main sequence galaxies at higher $z$ (out to $z \sim 2$) have warmer temperatures than those at $z = 0$. Note that this is not necessarily contradictory to the finding that $z \sim 2$ galaxies have colder SEDs than $z \sim 0$ galaxies of similar \lir; this is due to the dramatic evolution in the main sequence between these epochs.

\citet{magnelli14a} used \textit{Herschel} observations to propose stronger correlations of \tdust \ with specific star formation rate (sSFR) and parameterized distance to the main sequence (\Dms) rather than with \lir \ explicitly. The correlation is such that dust temperature is fixed for a fixed redshift and sSFR, implying that galaxies with a particular sSFR contain star-forming regions with similar $\langle\!U\!\rangle$. An increase in these star-forming regions results in the increase of SFR (star formation rate) with stellar mass (\mstar). Since starbursts have higher SFRs than galaxies with equivalent masses on the main sequence, more intense radiation fields coupled with higher densities could cause the elevated dust temperatures \citep{magnelli14a}.

A number of additional works also find an increasing dust temperature for main sequence galaxies, including \citet{bethermin17} and \citet{schreiber18}, among others. \citet{magdis12a} and \citet{bethermin15a} assert that the evolution of $\langle\!U\!\rangle$ points to an evolution of \tdust \ for all galaxies on the main sequence. \citet{magdis12a} and \citet{schreiber18} posit that the evolution of main sequence galaxy SEDs with redshift proves an evolution in \tdust. 

In contrast to the works that find an evolution toward hotter dust temperatures for main sequence galaxies at higher redshifts, several papers report that high redshift galaxies evolve toward {\it colder} temperatures at higher redshifts, in particular at $z \sim 2$ relative to galaxies at $z \sim 0$.  Such colder temperatures have been inferred based on galaxies of fixed SFR or \lir, and credited possibly to more extended dust geometries in high redshift galaxies \citep[e.g.][]{casey12b, symeonidis13a}. However, \citet{casey18a} argue that this perceived evolution toward colder temperatures could, in part, be a bias in underlying datasets that exist for $z \sim 0$ dust SEDs versus those at $z \sim 2$, and that most of the evolution, if it does exist, is between $0<z<0.4$. Further work is  needed to understand if this very low redshift  evolution is physically real or purely driven by limits in the existing datasets (Drew et al. in preparation).

Beyond $z \sim 0.5$ and out to $z \sim 5$, \citet{casey18b} find no evidence for redshift evolution of galaxies' SEDs in the \lir-\lpeak \ plane.  Similarly, \citet{dudzeviciute20a} find no evidence for evolution in dust temperature for galaxies as a function of fixed \lir. How can these seemingly disparate conclusions---that SEDs of main sequence galaxies evolve yet there is no observed evolution in \lir-\lpeak \ with $z$---be reconciled? These results are not, in fact, contradictory. In the main sequence, SFR (or \lir) evolves with $z$, but \lpeak \ does not evolve at fixed SFR. A direct correlation between \lir \ and \lpeak, or dust temperature, is well-established \citep[e.g.][]{sanders03, chapman03b, casey18b} and has been shown to be driven by increased dust heating in more luminous systems \citep{symeonidis13a} and to be independent of luminosity-limit selection effects \citep{sajina07a, lee13a}. As objects in the MS evolve toward higher luminosities (and correspondingly, higher $\langle\!U\!\rangle$) with $z$, \lpeak \ appears to evolve with redshift at fixed \mstar\ due to its correlation with \lir. Thus the perceived \tdust--$z$ evolution holds only for fixed sSFR, \Dms, or \mstar, and not for fixed SFR. 

In this paper, we explore the dependence of \lpeak\ on  SFR, sSFR, star formation surface density, and parameterized distance to the main sequence to better understand the underlying physical drivers of galaxies' integrated SED-averaged dust temperature.

\section{Sample and Observations} \label{sec:sample}
\subsection{Sample Selection}

Seven unlensed, spectroscopically-confirmed DSFGs were chosen from 400 arcmin$^2$
SCUBA-2 450\,\um \ and 850\,\um \ maps of the inner COSMOS field \citep{casey13a}. Of the 31 SCUBA-2 detected COSMOS sources with spectroscopic redshifts in \citet{casey17a}, we selected seven of the brightest sources (single-dish S$_{850} > 2$ mJy) to span a very broad range of dust SEDs from cold ($\sim$\,18\,K) to warm ($\sim$\,70\,K) based on their \textit{Herschel} SPIRE and SCUBA-2 photometry. The DSFGs were selected at the time for having known spectroscopic redshifts between $z = 1-3$ \citep{casey17a},  and updated information on this sample gives a total spanned redshift range of $1.4<z<4.6$. The galaxy initially thought to have the coldest SED, 450.03 (also known as AzTEC2), was identified initially to have a spectroscopic redshift of 1.123 from an OIR counterpart \citep{casey17a} but is now confirmed to sit at $z=4.61$ from a serendipitous detection of [CII] in our ALMA data and concurrent confirmation via CO(5-4) line emission from NOEMA observations \citep{jimenez20}. The ALMA centroid is offset from the OIR source by 1\farcs6 (see \citealt{casey17a} for details). The redshift for source 850.04 was originally estimated to be $z=1.436$, but closer inspection reveals a more likely redshift solution of $z=3.31^{+0.76}_{-0.81}$, which is a photometric redshift from the \citet{laigle16a} COSMOS catalog. This photometric counterpart is 1\farcs0 offset from the spectroscopically confirmed source at $z=1.436$ \citep{casey17a}  but significantly closer (0\farcs3) from the centroid of ALMA emission.

\subsection{ALMA Data}

\begin{figure}
\centering
\includegraphics[scale=.55]{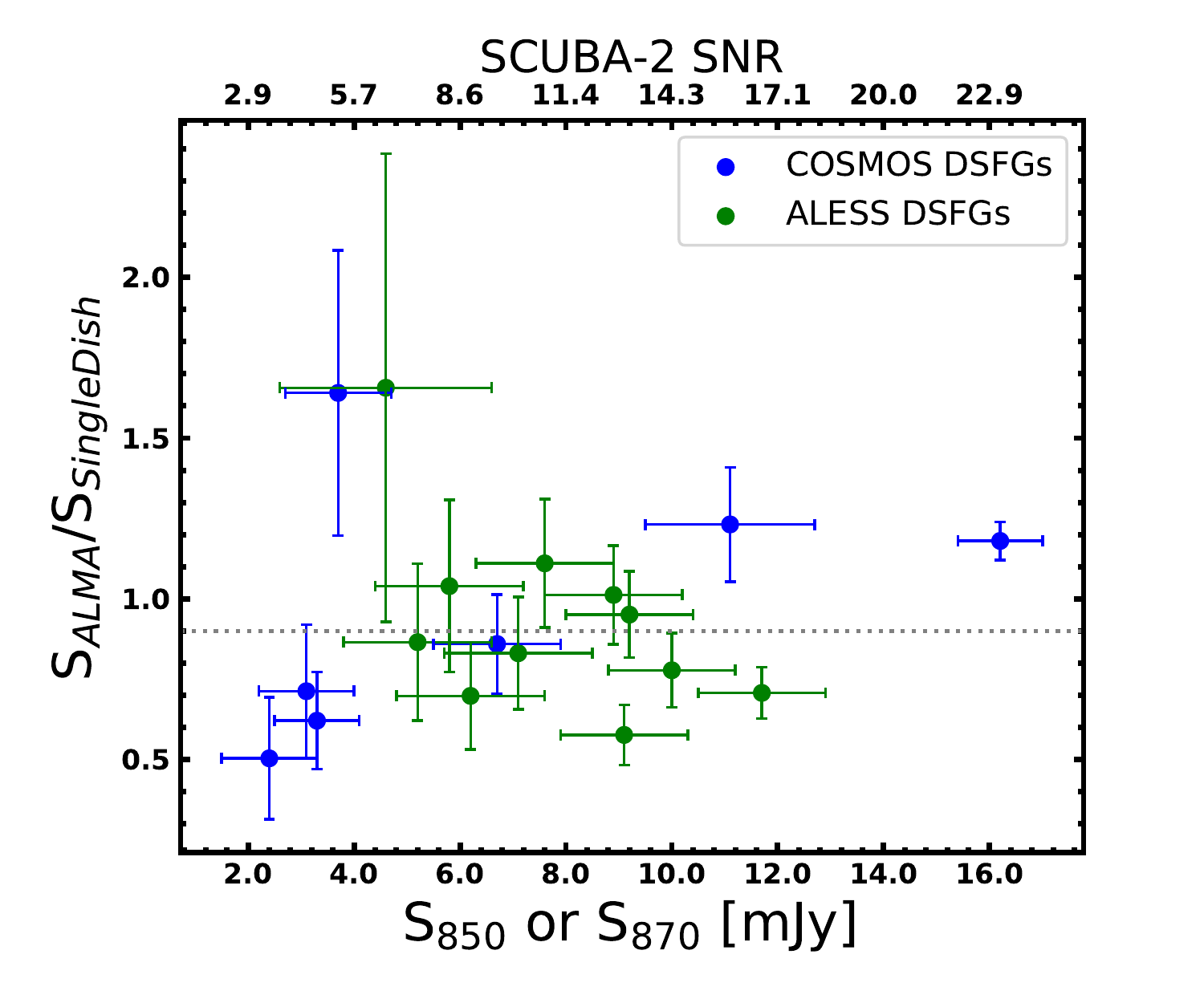}
\caption{Single-dish flux (S$_{850}$ or S$_{870}$) versus ratio of ALMA to single-dish flux (S$_{ALMA}$/S$_{SingleDish}$) for each galaxy. Single-dish fluxes are SCUBA-2 (850\,\um) for COSMOS sources and LABOCA (870\,\um) for ALESS sources. The SCUBA-2 SNR was calculated using $\sigma = 0.7$ mJy/beam. The dashed horizontal line at S$_{ALMA}$/S$_{SingleDish}$ = 0.9 indicates the expected flux ratio on the Rayleigh-Jeans tail of a blackbody. Three COSMOS sources (450.09, 450.03, and 850.04) have higher ALMA flux densities than SCUBA-2 flux densities, and the remaining COSMOS sources have lower ALMA flux densities. Four ALESS sources (15.1, 17.1, 45.1, and 112.1) have higher ALMA flux densities than LABOCA flux densities, and the remaining ALESS sources have lower ALMA flux densities.}
\label{fig:fluxplot}
\end{figure}

\begin{deluxetable*}{cccccccc}[htb]
\tablecaption{Redshifts, total flux densities, half-light radii as measured by \textsc{imfit} at 870\,\um, rest-frame SED peak wavelengths, infrared luminosities, star formation rates, and stellar masses for our  COSMOS SCUBA-2  and ALESS sources. Measurements of peak wavelengths, infrared luminosities, and star formation rates are done by the same method for both the COSMOS sample and the eleven ALESS sources we analyze. Measurements for 450.03 are for the total circularized size  and total \lir\ of both components. ALESS flux densities are from \citet{swinbank14a}, and sizes are from \citet{hodge16a} derived in a fully consistent manner as for our COSMOS sample. 
\label{tab:observations}}
\tablehead{
\colhead{Name} & \colhead{$z$} & \colhead{S$_{\rm 870}$} & \colhead{\re} & \colhead{\lpeak} & \colhead{\lir} & \colhead{SFR} & \colhead{M$_\star$} \\
\colhead{} & \colhead{} & \colhead{[mJy]} & \colhead{[\kpc]} & \colhead{[\um]} & \colhead{[\lsun]} & \colhead{[\msun/yr]} & \colhead{[\msun]} 
} 
\startdata
450.03 & 4.629\tablenotemark{a} & 19.12$\pm$0.16 & 1.35\tablenotemark{b}$\pm$0.06 & 71.4$\pm$7.8 & (3.28$\pm$0.68)\e{13} & 4900$\pm$1000 &  $-$\tablenotemark{c} \\
850.04 &  3.31 & 13.67$\pm$0.12 & 1.94$\pm$0.12 & 98$\pm$9   & (8.9$\pm$1.2)\e{12} & 1330 $\pm$180 &  (2.2$\pm$0.8)\e{10} \\
450.14 & 1.523 & 2.05$\pm$0.03 & 2.8$\pm$0.4    & 130$\pm$30 & (1.2$\pm$0.7)\e{12} & 190$\pm$100   &  $-$\tablenotemark{c} \\
850.60 & 1.457 & 2.21$\pm$0.04 & 1.81$\pm$0.28  & 75$\pm$9   & (1.9$\pm$ 0.4)\e{12} & 290$\pm$60   &  (2.1$^{+0.1}_{-0.2}$)\e{11} \\
850.95 & 1.556 & 1.21$\pm$0.03 & 6.1$\pm$1.4.   & 98$\pm$22  & (1.2$\pm$0.5)\e{12} & 180$\pm$80    &   (3.8$^{+0.4}_{-1.0}$)\e{10} \\
850.25 & 2.438 & 5.76$\pm$0.05 & 2.76$\pm$0.22  & 117$\pm$20 & (3.1$\pm$1.1)\e{12} & 460$\pm$170   &  (2.8$^{+0.1}_{-0.4}$)\e{11} \\
450.09 & 2.472 & 6.07$\pm$0.07 & 1.04$\pm$0.07  & 71$\pm$10  & (8.3$\pm$1.5)\e{12} & 1240$\pm$230  &  (2.2$^{+0.1}_{-0.4}$)\e{11}  \\
\hline
ALESS\,3.1   & 4.24 & 8.3$\pm$0.4 & 1.30$\pm$0.05 & 69$\pm$9     & (1.64$\pm$0.25)\e{13} & 2460$\pm$380 &  (2.3$^{+1.0}_{-1.6}$)\e{11} \\
ALESS\,5.1   & 2.86 & 7.8$\pm$0.7 & 2.00$\pm$0.10 & 106$\pm$12   & (4.9$\pm$1.0)\e{12}   & 730$\pm$150  &  (2.9$^{+0.2}_{-0.8}$)\e{11} \\
ALESS\,9.1   & 4.50 & 8.8$\pm$0.5 & 1.50$\pm$0.05 & 71$\pm$9     & (1.63$\pm$0.26)\e{13} & 2430$\pm$390 & (5.6$^{+1.3}_{-2.7}$)\e{11}  \\
ALESS\,10.1  & 0.76 & 5.3$\pm$0.5 & 2.60$\pm$0.20 & 190$\pm$9    & (2.3$\pm$0.4)\e{11}   & 34.5$\pm$5.5 &  (1.2$\pm$0.3)\e{10} \\
ALESS\,15.1  & 1.93 & 9.0$\pm$0.4 & 2.40$\pm$0.10 & 125$\pm$5    & (2.95$\pm$0.24)\e{12} & 440$\pm$40   &  (3.1$^{+1.3}_{-1.8}$)\e{11} \\
ALESS\,17.1  & 1.54 & 8.4$\pm$0.5 & 1.75$\pm$0.05 & 138$\pm$5    & (1.88$\pm$0.18)\e{12} & 281$\pm$27   &  (2.3$^{+0.1}_{-1.7}$)\e{11} \\
ALESS\,29.1  & 1.44 & 5.9$\pm$0.4 & 1.35$\pm$0.05 & 136$\pm$12   & (1.37$\pm$0.23)\e{12} & 200$\pm$30   &  (3.0$^{+1.2}_{-2.1}$)\e{11} \\
ALESS\,39.1  & 2.44 & 4.3$\pm$0.3 & 1.95$\pm$0.10 & 100$\pm$7    & (3.3$\pm$0.4)\e{12}   & 490$\pm$70   &  (4.5$^{+1.0}_{-1.6}$)\e{10} \\
ALESS\,45.1  & 2.34 & 6.0$\pm$0.5 & 2.15$\pm$0.15 & 117$\pm$10   & (2.7$\pm$0.5)\e{12}   & 400$\pm$70   &  (5.1$^{+1.5}_{-2.3}$)\e{11} \\
ALESS\,67.1  & 2.12 & 4.5$\pm$0.4 & 1.85$\pm$0.15 & 72.2$\pm$2.6 & (9.4$\pm$0.9)\e{12}   & 1400$\pm$140 &  (2.4$^{+2.4}_{-2.1}$)\e{11} \\
ALESS\,112.1 & 2.32 & 7.6$\pm$0.5 & 1.90$\pm$0.10 & 109$\pm$6    & (4.1$\pm$0.5)\e{12}   & 610$\pm$70   &  (1.0$^{+0.4}_{-0.5}$)\e{11} \\ 
\enddata
\tablenotetext{a}{The precision quoted on the redshift indicates whether or not the source has a spectroscopic redshift (four significant figures) or a photometric redshift (three significant figures).}
\tablenotetext{b}{Total circularized size of both components of 450.03. The individual half-light radii measured by  \textsc{imfit} are 1.24\,kpc for the brightest source and 1.08\,kpc for the secondary source.}
\tablenotetext{c}{Stellar masses are not provided for 450.03 or 450.14.  In the case of 450.03, this is due to significant blending with several foreground galaxies on angular scales that precludes measurement of near-infrared {\it Spitzer} flux densities; in the case of 450.14, there is a superposition of two galaxies of different redshifts within a 2$''$ region, making a proper separation of the background/foreground stellar mass distribution impossible.  Both cases are discussed at greater length in \citet{casey17a}.}
\end{deluxetable*}

Observations of these seven DSFGs were carried out as part of the ALMA Cycle 3 program 2015.1.00568.S (PI Casey). With a requested spatial resolution of 0\farcs1, observations were made in both compact (C36-3) and extended (C36-6) configurations. The compact and extended configurations had maximum baselines of 462.9\,m \ and 1.8\,km, respectively, with 36 antennas. The observations targeted dust continuum emission at the nominal band 7 frequency centered at 345 GHz or 870\,\um. At this frequency, ALMA has a 17\farcs3 primary beam (FWHM), and we utilized the “single continuum” spectral mode. 

We reduced and calibrated the raw data from the compact configuration to produce \textit{uv}-data products using the Common Astronomy Software Applications (\textsc{casa}) version 4.5.3. During calibration, data from antennas with irregular amplitudes were flagged. To produce optimum images for total flux recovery, we imaged the compact-configuration data using the \textsc{clean} algorithm of \textsc{casa} with Briggs weighting and robust = 2 (equivalent to natural weighting) and applied a primary beam correction. The resulting images have a resolution of 0\farcs68 $\times$ 0\farcs49. The compact data set was used in calculating the flux densities, which are given in Table \ref{tab:observations}. 

We created a higher resolution set of maps by combining the raw data from the compact and extended configurations, then reducing and calibrating the data using the same method described above. We tested different robust values with the \textsc{clean} algorithm and found the resulting size of each source to be consistent and independent of the robust parameter. Thus, we adopt a robust value of 1 for the full sample as it maximizes the source signal-to-noise for the most extended source in the sample (850.95) while maintaining a relative high angular resolution. The resulting resolution is 0\farcs24 $\times$ 0\farcs27. This does not meet the requested 0\farcs12 resolution, but even with a robust value of 0, the spatial resolution is still not as good as requested. The combined data was used to calculate the effective radii, which are also given in Table \ref{tab:observations}.  Images of the sample are shown in Figure \ref{fig:dustmass}  and overlays with optical/near-IR data are shown in Figure~\ref{fig:oircutouts} on the same scale. 

\subsection{Biases in the Sample?}
\label{sec:biases}
The data sample used in this analysis does not fully sample the full parent population of galaxy dust SEDs but is representative of the breadth of DSFGs at high-$z$. The COSMOS sources from \citet{casey17a} were chosen explicitly to have the broadest range of SEDs and are selected at either of 450\um \ or 850\um, which together is unbiased with respect to SED dust temperatures in the range of 20K--100K \citep[e.g.][]{roseboom13a,casey13a}. The ALESS sample selected at 870\um \ only is more likely to be biased toward colder dust temperatures \citep{chapman04c, casey09b}, but we see a large range of temperatures well represented in the sample, similar to the COSMOS sample. 
 Furthermore, while 870\um\ selection alone is intrinsically biased and favoring sources with colder SEDs, those that have well-sampled FIR SEDs (and high-resolution ALMA imaging) are not biased with respect to $\lambda_{\rm peak}$ because detection in the {\it Herschel} bands bracketing the SED peak is implicitly required. In other words, many ALESS sources without {\it Herschel} SPIRE detections (i.e. those with cold SEDs) would not have well-sampled FIR SEDs, are unlikely to have received high-resolution ALMA follow-up, and thus are excluded from our analysis here.

The sample (both COSMOS and ALESS DSFGs) could be seen as biased toward galaxies with high SFRs, but sources with high SFRs are the only ones for which the high spatial resolution measurements made in this paper are possible with a reasonable investment in ALMA time.   This has potential impact on our ability to probe the relationship between \lpeak\ and \lir\ in a fully unbaised way, given the limited dynamic range compared to galaxy populations at large.  This is discussed more fully in section~\ref{sec:lpeaklir}. The combined sample does sample galaxies' relative distance from the main sequence well, in line with the expected dynamic range of much larger populations of studied galaxies.  They span two orders in magnitude in specific star-formation rate from $\sim$0.6--60\,Gyr$^{-1}$, and about one order of magnitude in orthogonal distance from the main sequence; more methods to quantify galaxies' distance to the main sequence are discussed later in section~\ref{sec:mainsequence}.

The primary goal of this work is to assess the relative strength of correlation between luminosity-weighted dust temperature (as measured observationally through the proxy $\lambda_{\rm peak}$) and other physical characteristics of high-$z$ galaxies, including quantities constrained by the FIR-emitting region as measured by high-resolution ALMA dust continuum imaging.  Thus, this set of observations constitutes a well-constructed and relatively unbiased sample for our purposes.

\section{Calculations \& Results} \label{sec:results}

\subsection{ ALMA 870\um \ Flux Densities}\label{sec:fluxes}
	
Total integrated flux densities were measured directly from the compact ALMA data using a method similar to the one described in \citet{hodge16a}. We converted the maps from Jy beam$^{-1}$ to mJy pixel$^{-1}$ and masked emission less than 2$\sigma$. We then summed contiguous emission within an aperture of radius 3 $\times$ $b_{\rm maj}$, where $b_{\rm maj}$ is the FWHM (major axis) of the synthesized beam. The resulting total band 7 flux densities are given in Table \ref{tab:observations}. The median flux density of our seven sources is 5.8 $\pm$ 3.7 mJy at 870\um.

In general, when measuring the flux density, only contiguous emission was summed, with one exception. Source 450.03 is composed of two sources separated by 3\arcsec \ at the same redshift \citep{jimenez20}, both of which were included in the total reported flux density. 

To check our flux density calculations, we used the \textsc{imfit} tool in \textsc{casa} and found that our calculations agree with the \textsc{imfit}-derived densities for all but one of our sources. Source 850.95 differed with less than 2$\sigma$ significance of deviation. The average ratio between \textsc{imfit} and aperture flux density measurements was $0.96\pm0.10$.

In addition, we compare the flux densities as measured by ALMA to the single-dish flux densities obtained for the COSMOS sources from SCUBA-2 at 850\,\um\ using the single-dish JCMT and for the ALESS sources from LABOCA at 870\,\um. The average ratio of ALMA to single-dish flux densities was $0.94\pm0.21$, as shown in Figure \ref{fig:fluxplot}. Note that the expected ratio of 850\,\um \ to 870\,\um \ flux is $\approx$ 0.9 for Rayleigh-Jeans tail emission.

\subsection{Measuring Dust Sizes and Morphology}\label{sec:sizes}

We quantitatively examine the size of the 870\um\ dust emission by fitting each source in the image plane with two-dimensional Sersic and Gaussian models. The Sersic model was applied using \textsc{galfit}, and the Gaussian model was applied using \textsc{casa} \textsc{imfit}. 

Dust emission half-light effective radii (\re) were measured using \textsc{galfit} on the ALMA maps in the image plane. \textsc{galfit} is a 2D fitting tool that models the profiles of astronomical objects using parametric functions \citep{peng10a}. Here we use \textsc{galfit} on ALMA data to test for non-Gaussian source morphologies, which is most easily applied in the image plane. The full width at half maximum (FWHM) was measured for each source using \textsc{casa} \textsc{imfit}, which is also applied in the image plane. Most of our sources are moderately resolved---more than two beams across in size---so size measurements are unlikely to be affected by fitting methods. Thus, making measurements in the image plane rather than the \textit{uv} plane is justified.

\begin{figure*}
\centering
\includegraphics{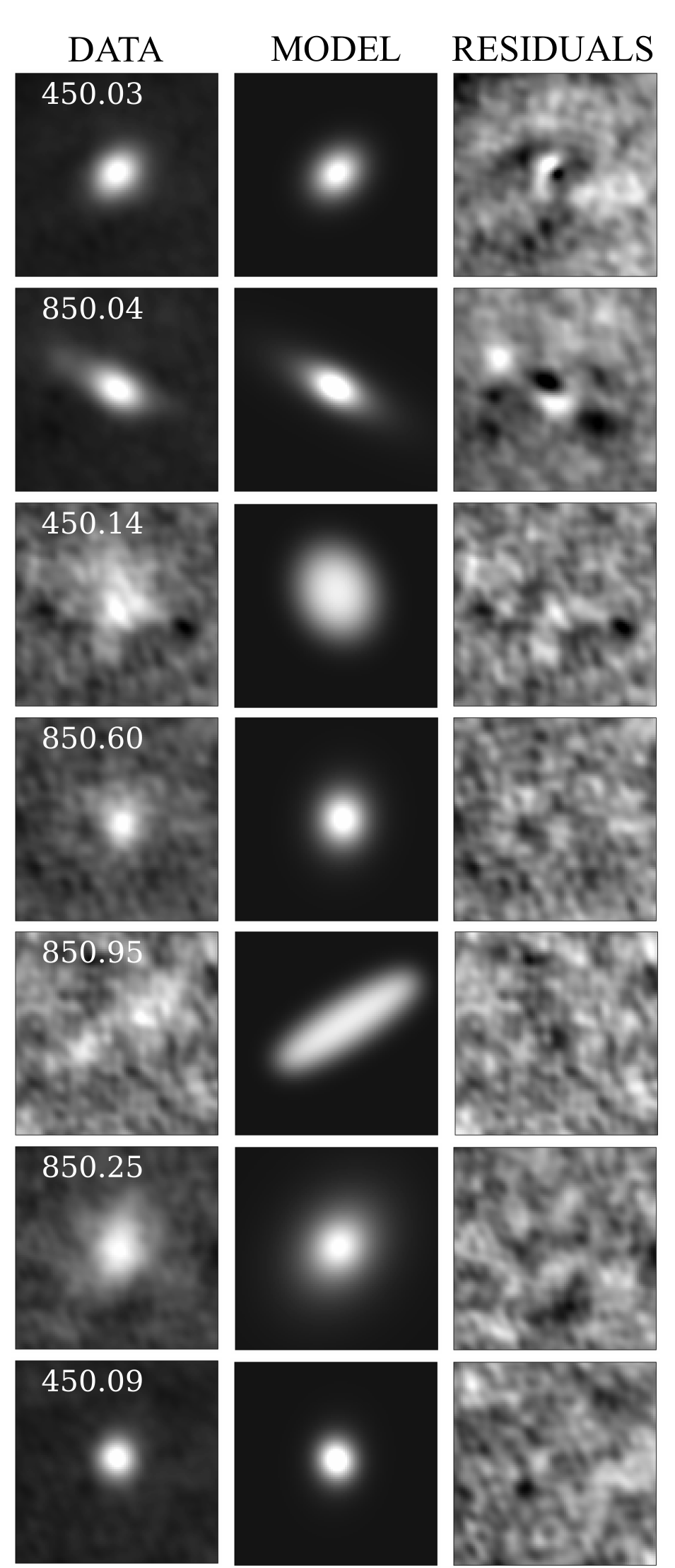}
\caption{2\arcsec$\times$2\arcsec model and residual images generated by GALFIT for each source. Images on the far left are the original ALMA 870\,\um \ images with robust = 1. Only the brightest component of 450.03 is shown.}
\label{fig:galfit}
\end{figure*}

To address whether the weighting applied with the \textsc{clean} algorithm has an impact on size measurement in the image plane, measurements were also made in the \textit{uv} plane using \textsc{casa} \textsc{uvmodelfit} to fit a Gaussian model. The \textsc{uvmodelfit} and \textsc{imfit} measurements agree within uncertainties for four of our sources. There is a discrepancy between \textit{uv} and image plane sizes for 450.03 and 850.25, which we attribute to significant excess emission from the core of the systems, which is evident in the \textsc{imfit} residual images. There was also a discrepancy in the measurements for source 850.95, which we attribute to the \textsc{imfit} model extending beyond the region of significant flux associated with the source. In general, the measured Gaussian FWHMs in the \textit{uv} plane are systematically offset from the FWHMs in the image plane measured by \textsc{imfit} such that the median ratio of \textit{uv}-plane to  image-plane FWHMs is 0.81 $\pm$ 0.08. Our scientific results rely on relative size comparisons of objects within the sample (and extended to the ALESS sample) rather than absolute sizes. 
The results of \textsc{galfit} are shown in Figure \ref{fig:galfit} along with the residuals and models. It is a known problem that errors from \textsc{galfit} are underestimated; therefore, we used the \textsc{imfit} results in our analysis and report \textsc{imfit} sizes in Table \ref{tab:observations}.

The sizes measured here are the mass-weighted sizes--measured on the Rayleigh-Jeans tail of the SED--rather than the luminosity-weighted sizes, which we would measure closer to the peak of the SED. The only exceptions to this are those galaxies that sit at $z>3$ in our sample (5 of 18); at these redshifts, observed 870\um \ probes rest-frame wavelengths shortward of 200\um, which may be more prone to tracing an optically thick regime of the SED in some cases (dependent on the actual dust column). We explore the impact of these higher redshift sources on our final results later in Section \ref{sec:discussion}. A dearth of data in the literature makes it difficult to know whether measured size varies substantially along the RJ tail, but the limited one-off measurements that do exist suggest it is safe to assume any measurement on the tail should be consistent within measurement uncertainties. In addition, these sizes are distinct from the sizes of the galaxies traced by stellar emission; the two are often spatially distinct in DSFGs, such that the rest-frame UV/optical emission is heavily obscured by dust, and thus, does not serve as an adequate probe of the characteristic scale of the ISM \citep{biggs08a, chen15a, hodge15, hodge16a, lang19}.   It should also be noted that dust continuum sizes are systematically smaller and not entirely consistent with radio continuum sizes (which probe synchrotron emission scattering off of supernova remnants), the measurements of which exist for much larger samples than those with dust continuum size measurements \citep{miettinen17a}.

The effective radii measured by \textsc{galfit} range from 0\farcs13$-$0\farcs69 with a median of \re $=$ 0\farcs32 $\pm$ 0\farcs09. Using the redshift of each source, we convert these angular sizes to physical sizes, which range from 1.1$-$5.9 \kpc \ (median \re $= 2.7 \pm$ $0.8$\,kpc). Source 850.95 was an outlier at $5.9 \pm 3.3$ \kpc. It is known to be somewhat abnormally large as also traced by H$\alpha$ kinematics \citep{drew18}. Source 850.95 also has the lowest signal-to-noise ratio among our sources. The remaining sources ranged from 1.1$-$3.2\,\kpc \ (median \re $= 2.3 \pm$ 0.5\,\kpc). 
 
The FWHMs (deconvolved from the restoring beam) measured by \textsc{imfit} are shown in Table \ref{tab:observations}. They range from 0\farcs26$-$1\farcs45, which convert to 2.1$-$12.3\,\kpc \ on physical scales. The half light radii quoted in Table \ref{tab:observations} are equivalent to these FWHM measurements divided by two. The median Gaussian profile has a FWHM $= 3.9 \pm 1.6$\ kpc. Excluding 850.95, the median is   $3.8 \pm 1.4$\ kpc. 

The Gaussian profiles of the ALESS sources included in our analysis were measured via the same technique used here \citep{hodge16a}, where the median FWHM $= 3.8 \pm 0.5$\,kpc. The median size of our sources agrees with that of the ALESS sources, despite the fact that the ALESS sample is  brighter than all but two of our sources. Overall, the DSFG population from which the ALESS and COSMOS SCUBA-2 DSFGs are drawn are very similar, with the exception of the ALESS median total 870 \um\ flux densities being $1.3\times$ brighter. The median of our sources' half-light radii is also consistent within uncertainty with that found by \citet{spilker16} and \citet{gullberg19}.

The Sersic indices of our sources range from $n=$0.3 to 2.1, with a median of $n = 0.8 \pm 0.4$. Our median $n$ is consistent with that of the ALESS sources in \citet{hodge16a}, whose median Sersic index $n = 0.9 \pm 0.2$ implies a non-Gaussian, exponential disk morphology. \citet{gullberg19} also found an exponential disk emission profile for a sample of DSFGs. 

\subsection{SED Integrated Dust Temperature}\label{sec:seds}

Wien's Law states that \lpeak $=$ $b$/\tdust \ (where $b$ $=$ {2.898\e{-3}} m\,K) for an idealized blackbody, but this relationship deviates depending on the adopted opacity model of the blackbody as shown in Figure 20 of \citet*{casey14a}. Thus, we parameterize the luminosity-weighted dust temperature (\tdust) instead as the observable \lpeak, the rest-frame wavelength where the dust spectral energy distribution (SED) peaks. The observable $\lambda_{\rm peak}$ is not sensitive to opacity assumptions and is a better trace of the observed peak in the SED in any observational sample without high quality constraints on the underlying dust opacity.

SEDs in this work were fitted using the technique described in \citet{casey12a}, based on a single dust temperature modified blackbody fit with a mid-infrared power law shortward of $\sim$\,70\,\um \ in the rest frame. The mid-infrared powerlaw represents a superposition of modified blackbodies at warm to hot temperatures, where the least amount of dust mass is heated to the hottest temperatures, and the vast majority of the mass of dust in the ISM is at much colder temperatures, represented by the temperature of the fit at rest-frame wavelengths $\simgt$200\um.  Because our sample has somewhat limited rest-frame mid-infrared photometric constraints, we fix the mid-infrared power law slope to $\alpha_{\rm MIR}=2$, representative of the average measured power law slope for local LIRGs and ULIRGs that have well-sampled mid-IR SEDs \citep[e.g.][]{u12a}.

The photometric points used in the fitting include ALMA (870\,\um), Herschel SPIRE (250, 350, 500\,\um), SCUBA-2 (450, 850\,\um), and Spitzer (24\,\um).   Points are weighted relative to their signal-to-noise, with exception of the {\it Spitzer} 24\,\um\ point whose signal-to-noise is capped at 10 for the purposes of these bulk SED fits.  The 24\,\um\ photometry often is a very high signal-to-noise measurement but is not always in agreement with the underlying SED model for dust continuum emission alone, due to the possible contribution of PAH emission features and silicate absorption in the rest-frame mid-infrared but not directly modeled here. We also note that use of the response function of the various far-infrared/millimeter passbands is not necessary using the \citet{casey12a} fitting method given the lack of  $\approx$1\% precision in the photometry in   the FIR regime \citep[see also][]{casey20a}. 

\begin{figure*}
\centering
\includegraphics[width=1.99\columnwidth]{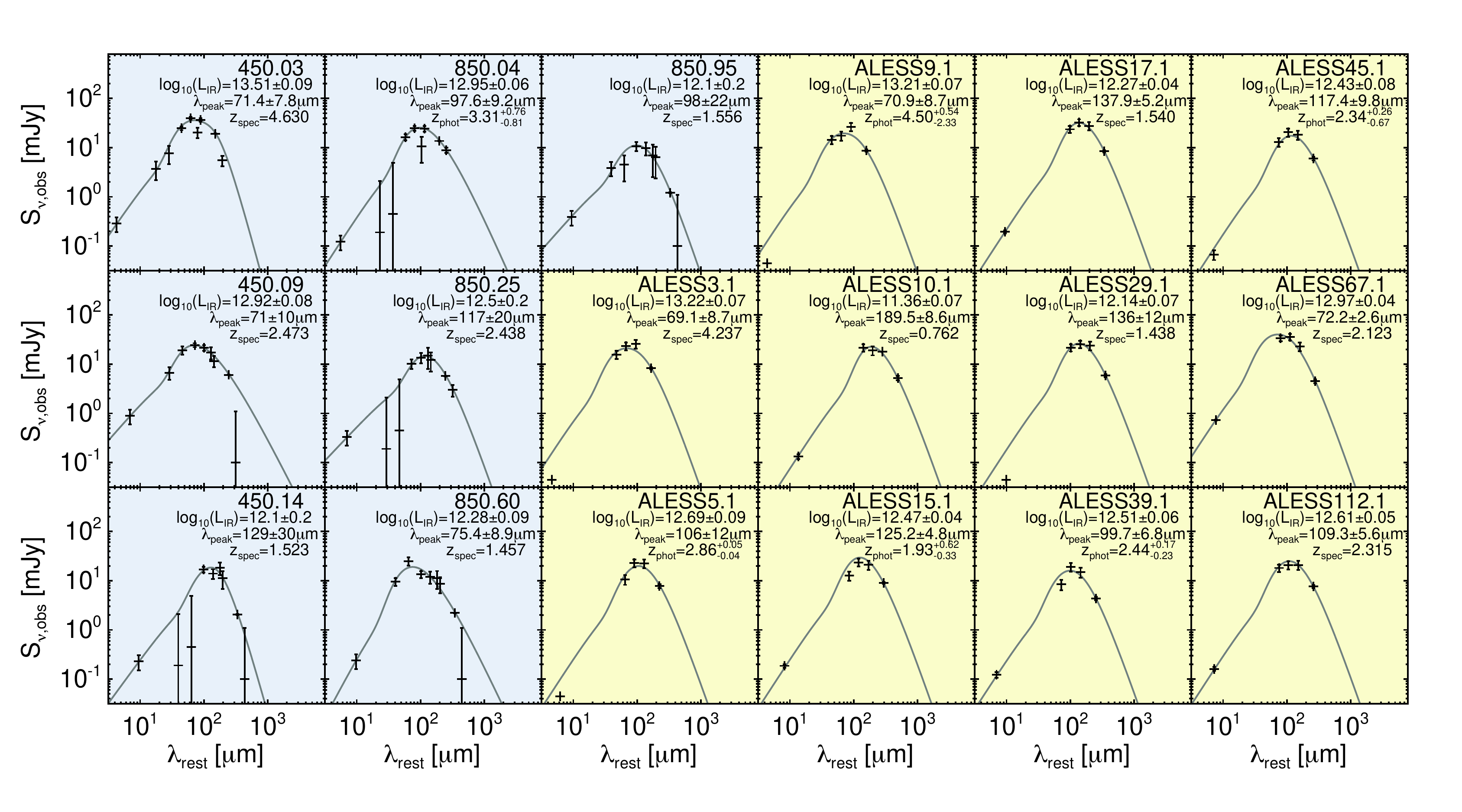}
\caption{Infrared SEDs for our sample and the ALESS sample included in our analysis. The photometry is as reported in \citet{casey13a,casey17a}, \citet{swinbank14a}, and \citet{hodge16a}, and SEDs are fit using the single modified blackbody plus mid-infrared powerlaw technique detailed in \citet{casey12a}. We tested different permutations of SED fits (fixing or allowing variance in the mid-infrared powerlaw slope, the emissivity spectral index, or assumed opacity of the dust near the peak) and find that measurement of \lir \ or \lpeak \ are insensitive to higher-order permutations within the margin of error. Noted in each panel is the source redshift, \lir, and \lpeak \ as also stated in Table \ref{tab:observations}.}
\label{fig:SEDs}
\end{figure*}

 While the bulk SED, including the measured IR luminosity or rest-frame peak wavelength, is not impacted by the adopted opacity model, here we adopt a fixed opacity model for all SEDs such that $\tau$=1 at 200\,\um \ following \citet{conley11a}.  Given that our observations include spatially-resolved dust maps, we can test if this assumption is appropriate for this sample by inferring the wavelengths at which $\tau$=1 from the dust column density.  Using the total 870\um\ flux densities from section~\ref{sec:fluxes}, sizes from section~\ref{sec:sizes}, a fixed mass-weighted dust temperature of 25\,K, and dust mass absorption coefficients from \citet{li01a}, we infer that the wavelength at which the SEDs become optically thick to vary from 10--300\,\um\ (with an average $\lambda(\tau=1)=80\pm50$\,\um).  Given the uncertainties in the derivation of dust mass itself, we choose to apply the uniform assumption to our SEDs rather than tuning each fit according to its measured dust column. The SEDs are shown in Figure \ref{fig:SEDs}, and the measured peak wavelengths and IR luminosities are also given in Table \ref{tab:observations}. We also re-fit SEDs for the ALESS sources described in \citet{hodge16a} whose multiwavelength photometry is given in \citet{swinbank14a}. Fitting SEDs for both the COSMOS SCUBA-2 DSFGs and the ALESS sources was done in a fully consistent manner.

\subsection{Star Formation Rates and Stellar Masses}\label{sec:mainsequence}

Star formation rates (SFRs) were calculated using the total infrared luminosities derived from the SED fits in section~\ref{sec:seds} and the logarithmic constant of conversion, $\rm \log(C_{IR}) = 43.41$ from \citet{kennicutt12}. We do not have reliable stellar mass estimates for 450.03 and 450.14 due to foreground-background source contamination (see \citealt{casey17a}), so they are omitted from all analysis requiring  measurements of \mstar.  The stellar masses of the COSMOS sample are presented in  \citet{casey17a}, and for the ALESS sources  they are presented in \citet{dacunha15a}.
All stellar masses used in this study are derived with MAGPHYS \citep{dacunha15a, dacunha08} in a uniform fashion using stellar population models from \citet{bruzual03}. For the two sources for which the redshifts were corrected, we reran the MAGPHYS modeling accordingly. Note that the stellar masses of DSFGs are often highly uncertain, though not necessarily systematically so, as they can also be overestimated by applying an inappropriate star formation history \citep[see discussions in][]{hainline11a, michalowski12}. The uncertainties on the stellar masses are consistent with the broad uncertainties on DSFG stellar masses in general. We note that our MAGPHYS fits are used exclusively to derive stellar masses, and the MAGPHYS fits are not used in this work to infer FIR SED characteristics like dust temperature or SFR; this is because of the potential for the rest-frame optical/near-infrared SED to be highly decoupled from FIR emission in DSFGs, and our desire to directly test the FIR fits alone and independently of the rest-frame optical/near-infrared emission used to measure the stellar masses. We also calculated the specific star formation rate (sSFR), where {\rm sSFR} = {\rm SFR}/\mstar.

\begin{figure}
\centering
\includegraphics[scale=.6]{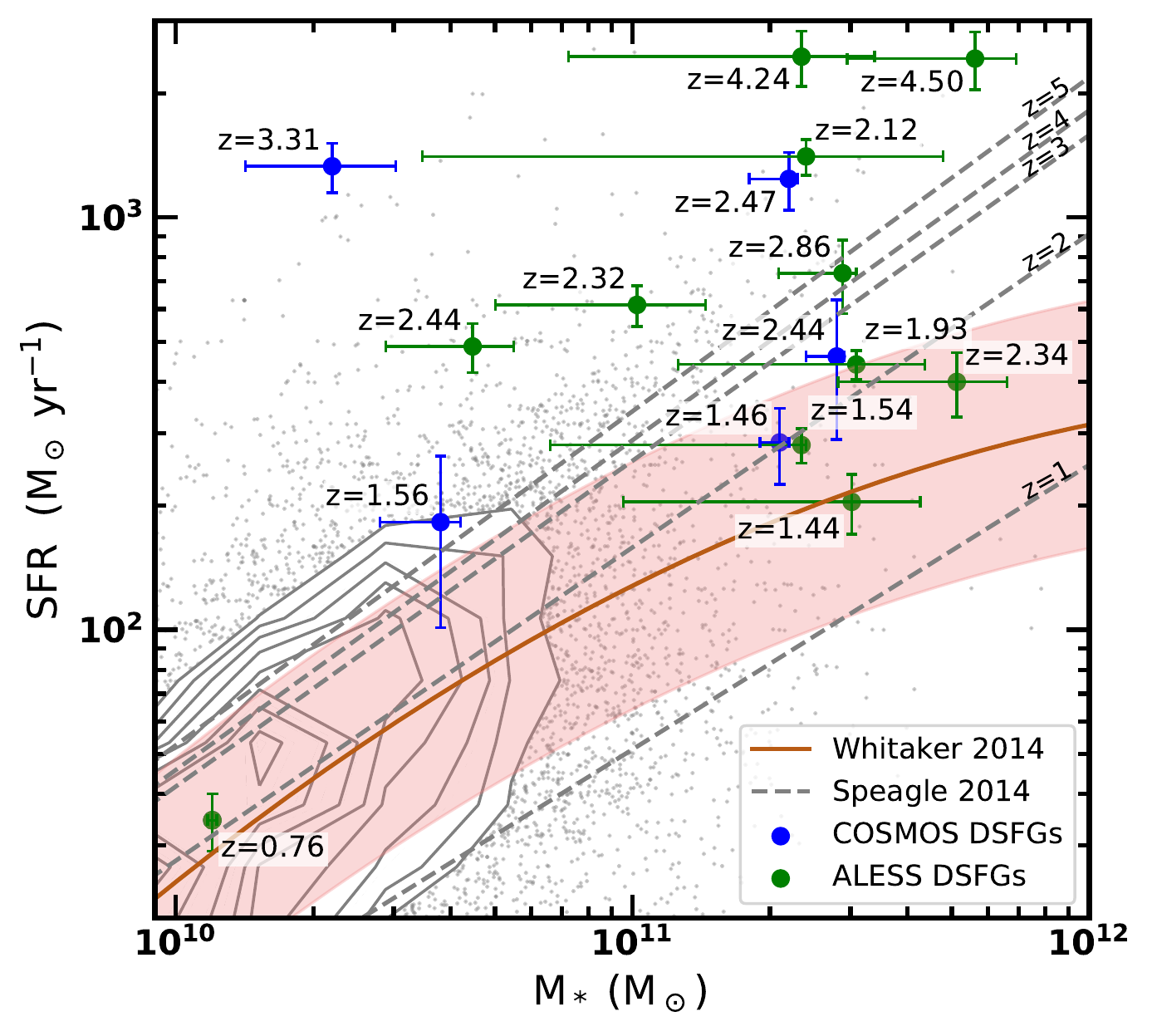}
\caption{Five of our COSMOS sources (blue) and eleven of the ALESS sources (green) overplotted on the galaxy main sequence. Two of our COSMOS sources (450.03 and 450.14) are not included here because we do not have reliable stellar mass estimates for them. The orange line is the best-fit polynomial of the form used in \citet{whitaker14a} at $z$ = 2. Note that there is significant scatter in the main sequence, and the average scatter is represented here as the red shaded region, with a width of 0.3 dex. Sources in the red shaded region or with error bars extending into the red shaded region lie on the main sequence. The dashed lines are the cosmic time-dependent best fits from \citet{speagle14a} for redshifts $z$ = 1, 2, 3, 4, and 5.}  
\label{fig:mainsequence}
\end{figure}

Figure \ref{fig:mainsequence} shows the COSMOS and ALESS sources in relation to the SFR--\mstar \ relation, also called the galaxy main sequence. In red, we show the main sequence of star-forming galaxies with a quadratic function of the form used in \citet{whitaker14a} applied to COSMOS data (1.5 $\leq$ z $\leq$ 2.5) from \citet{laigle16a}. We applied a Markov Chain Monte Carlo (MCMC) approach to derive the parameters of the fit for the broader COSMOS data set. Also shown also are the cosmic time-dependent best fits from \citet{speagle14a} for redshifts $z =$ 1, 2, 3, 4, and 5. 

 For each galaxy in our sample, we also compute a parameterized `distance' to the main sequence, or \Dms.  \Dms\ represents the orthogonal distance between the main sequence at the source's redshift and its measured value in the $\log$\mstar--$\log$SFR plane.  We compute orthogonal distances to the main sequence rather than projected distances in SFR or sSFR to minimize the impact of measurement uncertainties in both \mstar \ and SFR for each galaxy.  The main sequence used to calculate \Dms\ for each source were the cosmic time-dependent curves given in \citet{speagle14a}; however, we note that computing \Dms\ using a quadratic form of the main sequence as in \citet{whitaker14a} produces similar results, ultimately not changing the conclusions of our work no matter which definition of \Dms\ is adopted.  Similarly, as we discuss later in section~\ref{sec:discussion}, we test for correlation between \lpeak\ and $\Delta$sSFR, finding results that are consistent with our adopted \Dms.

\section{Discussion} \label{sec:discussion} 

\subsection{Expected Relationship of Dust Geometry to Dust Temperature}

Most galaxies for which dust temperatures have been constrained lack spatially-resolved measurements needed to investigate the effects of dust geometry on dust temperature. Thus, the underlying physical driver of a galaxy's globally-averaged dust temperature has not been directly constrained. Before ALMA, there were small samples of high-$z$ galaxies that had resolved radio or millimeter sizes \citep{chapman04b, biggs08a, casey09b, younger09a}, but the observations needed to determine accurate sizes for larger samples of DSFGs require the order of magnitude improvements in sensitivity brought by ALMA \citep[see ][]{hodge16a,rujopakarn16a,simpson17,gullberg19a,elbaz18a}. In this paper, we have presented a sample of DSFGs with both high-resolution dust size constraints and good dust SED constraints, allowing a more thorough investigation into the physical drivers of galaxies' dust SEDs.

This study is theoretically motivated by the application of the Stefan-Boltzmann law on galaxy scales. Specifically, it relates the emergent IR luminosity to the effective size of the blackbody (\re) \ and temperature (\tdust) of an optically thick  spherical blackbody: 

\begin{equation}
\mathrm{L_{\rm IR} = 4\pi R_{e}^{2}\sigma T^{4}}
\label{eq:sf}
\end{equation}
\\
The Stefan-Boltzmann constant is $\sigma$ = 5.670\e{-5} \,erg\,s$^{-1}$\,cm$^{-2}$\,K$^{-4}$. Galaxies are certainly not as simple as stars in the application of such a relation.   Nevertheless, approximating galaxies' IR emission as an optically thick shell of dust surrounding an obscured point source (i.e. deeply embedded OB associations in star-forming regions) would result in the following expected relation between the dust's temperature, IR luminosity and dust shell radius:
\begin{equation}
    \log\Sigma_{\rm IR} = \log(4\sigma) + 4\log(T_{\rm dust})
    \label{eq:sphereicalSB}
\end{equation}
Where $\Sigma_{\rm IR}$ is the IR luminosity surface density, =$L_{\rm IR}/\pi R_{e}^2$.  Of course, galaxies' geometry is unlikely to be well represented as a spherical shell, and the surface area of the emergent IR flux does indeed play a role in the expected relation.  If instead of a spherical shell we assume some uniform planar distribution of UV emission underneath an optically thick screen (as in a galaxy disk, with half-light radius $R_{e}$) we would expect the following relationship instead:
\begin{equation}
    \log\Sigma_{\rm IR} = \log(\sigma) + 4\log(T_{\rm dust})
    \label{eq:sigmairtdust}
\end{equation}
which is a factor of four different than the spherical shell: one factor of two given the modified surface area of a disk as 2$\pi R_{e}^2$ and one factor of two accounting for $R_{e}^2$ being a measured {\it half-light} radius.  Then given our use of optically thick SEDs (described in section~\ref{sec:seds}), where the relationship between dust temperature and \lpeak\ can be approximated as
\begin{equation}
\label{eq:lpeaktd}
    \log(T_{\rm dust}/K) \approx 3.756-1.048\log(\lambda_{\rm peak}/\mu\!m)
\end{equation}
in the 15--70\,K range\footnote{Note that this only deviates by 5\%\ from Wien's law for optically thin, idealized blackbodies.}.  This gives a predicted relationship between \lpeak\ and $\Sigma_{\rm IR}$ of
\begin{equation}
    \log(\Sigma_{\rm IR}/[L_\odot\,{\rm kpc}^{-2}]) \approx 20.18 - 4.19\log(\lambda_{\rm peak}/\mu\!m)
    \label{eq:expected_lpeaksigmaIR}
\end{equation}
Note that expectation using Wien's Law directly instead of the approximation given in Eq~\ref{eq:lpeaktd} would yield
\begin{equation}
\begin{array}{ll}
    \log(\Sigma_{\rm IR}/[L_\odot\,{\rm kpc}^{-2}])& = \log(\sigma b^4)-4\log(\lambda_{\rm peak}/\mu\!m)\\
    & \approx 19.00 - 4\log(\lambda_{\rm peak}/\mu\!m)\\
    \end{array}
    \label{eq:wienslaw_lpeaksigmaIR}
\end{equation}
where $b$=2.898\e{3}\,\um\,K.

One key caveat here is that the Stefan-Boltzmann Law should only be applicable to optically thick blackbodies, and the dust in galaxies' ISM is often far from optically thick at all wavelengths (of course, it is a huge benefit that ISM dust is often optically thin on the Rayleigh-Jeans tail of blackbody emission, allowing for a direct measurement of galaxies' dust mass from measured flux density). We return to the conundrum of dust opacity again in section~\ref{sec:lpeaksigmair}, where we derive the relationship between $\Sigma_{\rm IR}$ and \lpeak\ empirically.

\subsection{What correlates best with dust temperature?}

Here we investigate the relative strengths of correlations between \lpeak\ (our observational proxy for dust temperature) and each of the following physical properties of galaxies: sSFR, \Dms,
SFR, and $\Sigma_{\rm IR}$. All of these quantities have been argued to correlate tightly with galaxies' dust temperatures, for example: sSFR and \Dms\ \citep[e.g.][]{magnelli14a}, SFR or L$_{\rm IR}$ \citep[e.g.][]{chapman03a,symeonidis13a}, and $\Sigma_{\rm IR}$ \citep[e.g.][]{chanial07,lutz16a}. For example,
\citet{magnelli14a} report that \tdust \ correlates more strongly with sSFR and parameterized distance to the main sequence (\Dms) than \lir \ (they lacked direct measures of \re \ so could not investigate \sfrsurfdens). In addition, they deduced that cold galaxies ($\lesssim$ 25K) sit on the main sequence and warm galaxies ($\gtrsim$ 30-80K) lie above the main sequence \citep{magnelli14a}.

\begin{figure*}
\centering
\includegraphics[scale=.5]{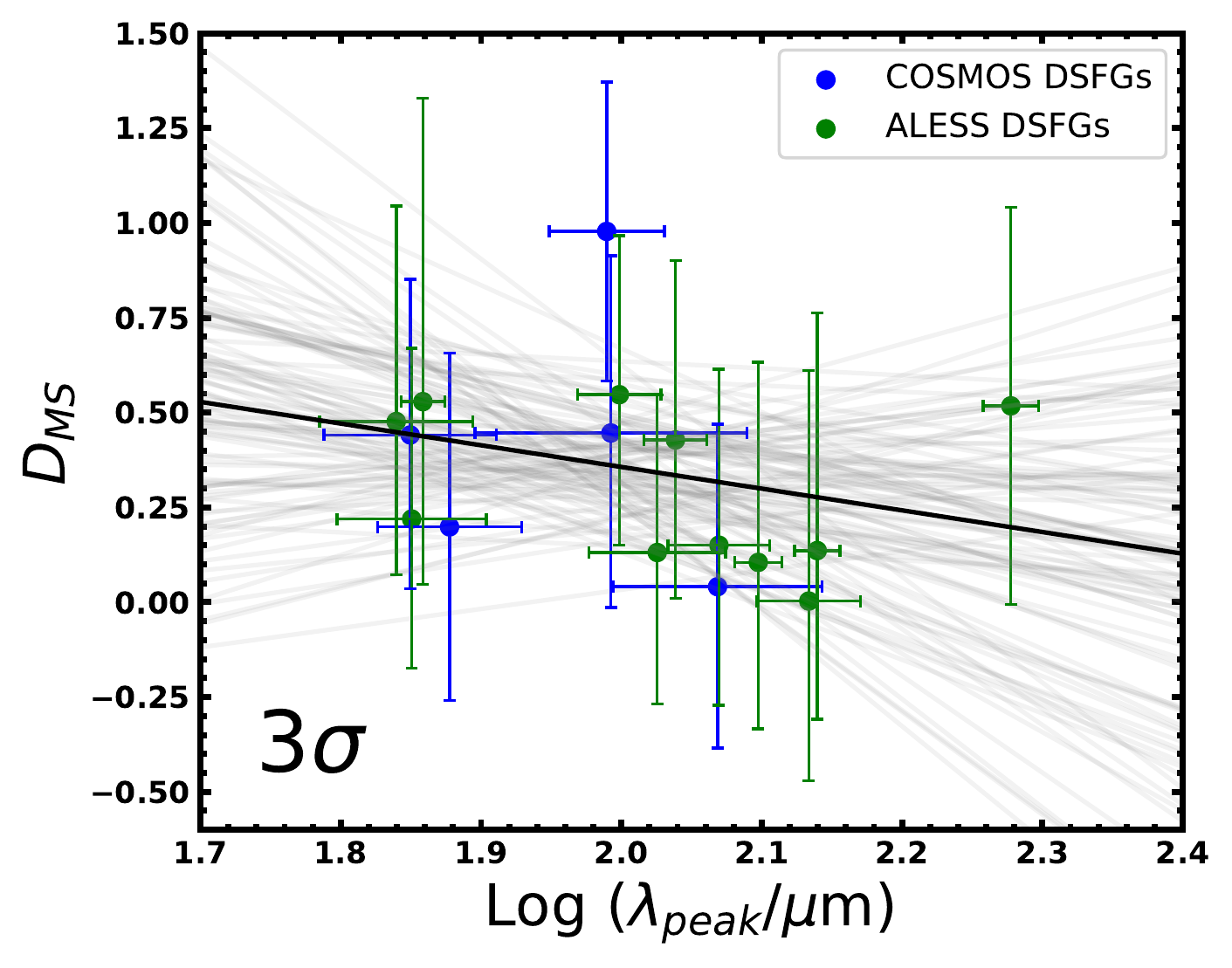}
\includegraphics[scale=.5]{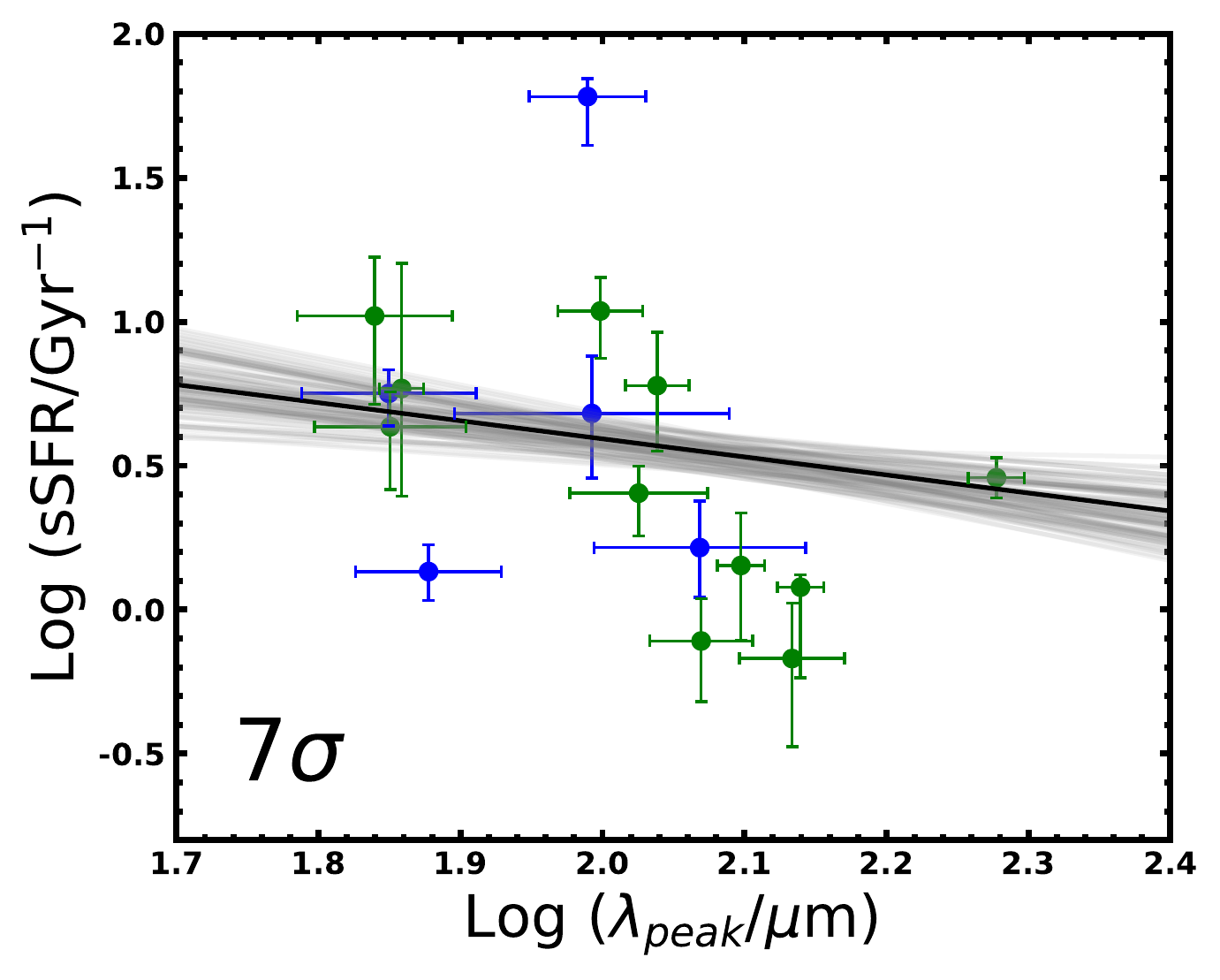}
\\
\includegraphics[scale=.5]{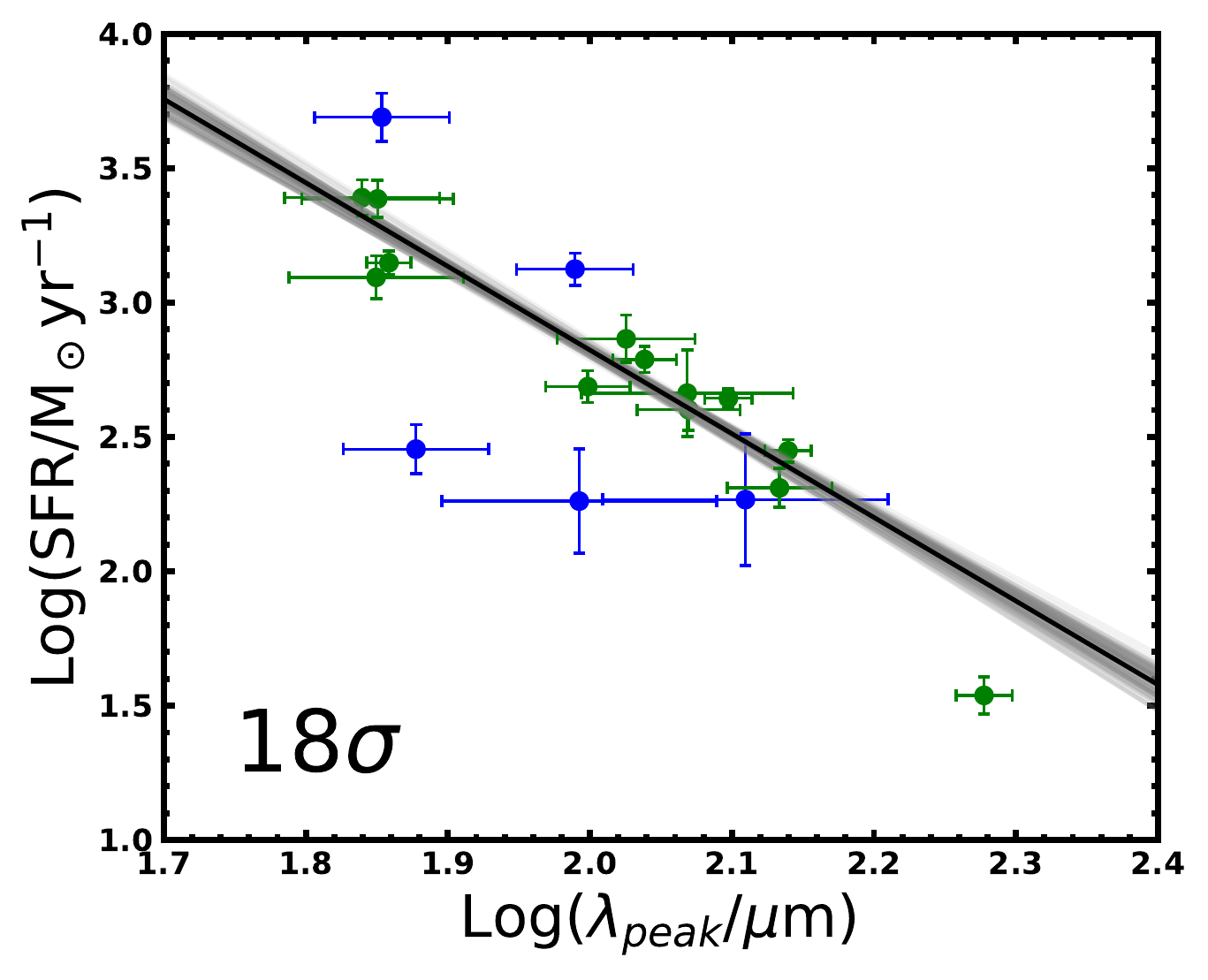}\includegraphics[scale=.5]{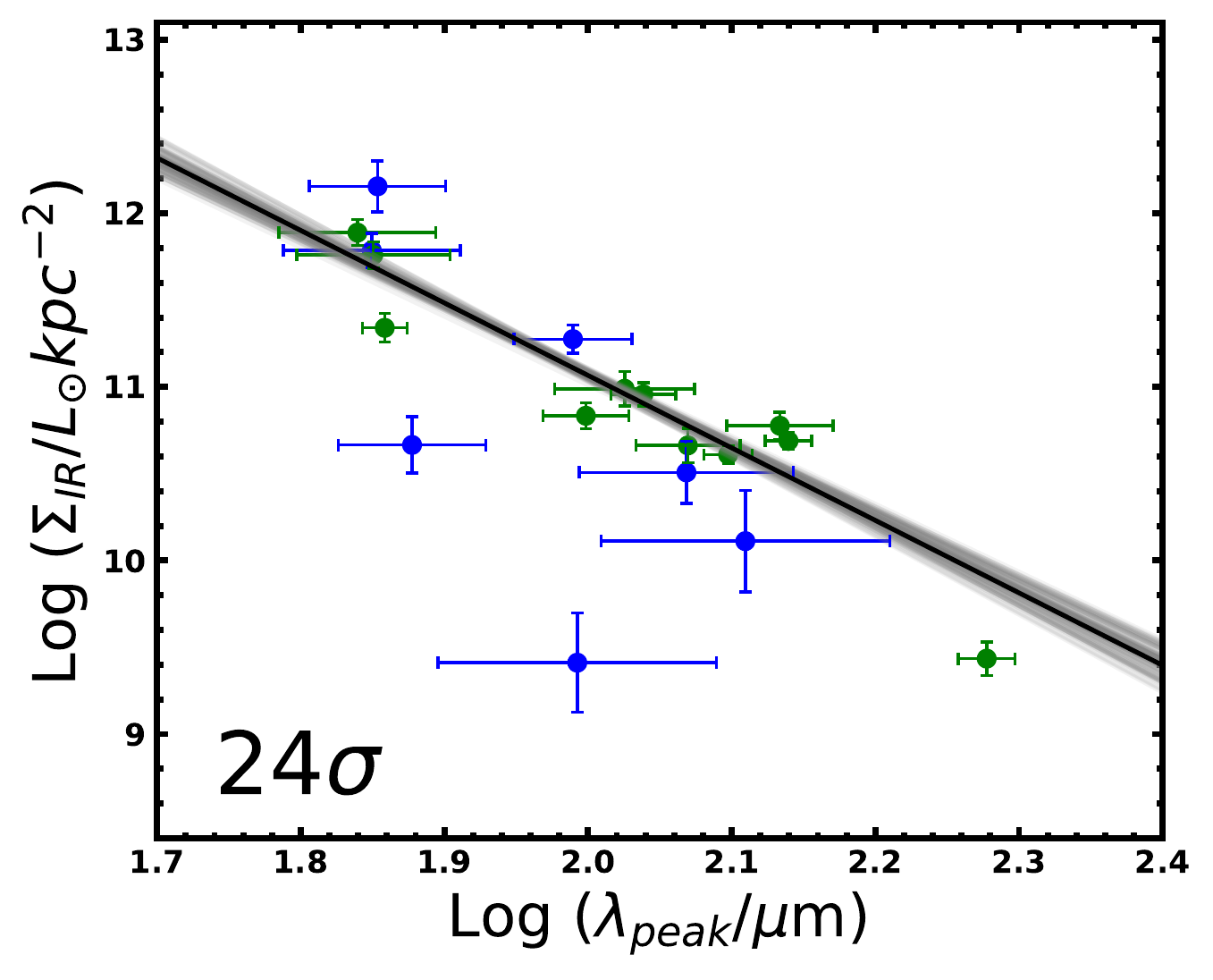}
\caption{Correlations between \lpeak \ and the parameterized distance to the galaxy main sequence, sSFR, SFR, and $\Sigma_{\rm IR}$. A linear model has been fitted in $\log-\log$ space using MCMC techniques. A representative sample of MCMC trial fits are shown in gray while the black line shows the best fit. The number in the lower left-hand corner of each plot represents the significance of the relation from a horizontal line, which would indicate no correlation. Larger numbers represent stronger correlations. We infer weak correlations of \lpeak \ with \Dms \ and sSFR and strong correlations of \lpeak \ with SFR and star formation surface density. }
\label{fig:correlations}
\end{figure*}

 Here we use this high-$z$ sample with well measured sizes and SEDs to directly constrain which of these physical correlations is tightest, and may therefore be the most fundamental to the internal ISM.  This may be of particular use to the community when only some of these quantities can be measured (for example, galaxies lacking high-resolution dust continuum imaging). We calculate the significance of each correlation as the deviation of the model from  no correlation, or a horizontal line in the given parameter space. The results are shown in Figure \ref{fig:correlations}. 

 We find that the  \lpeak \  vs. \Dms \ and  \lpeak\ vs. sSFR relationships correlate with 3$\sigma$ and 7$\sigma$ significance, respectively.  In contrast, the  \lpeak\ vs. SFR and  \lpeak\ vs. $\Sigma_{\rm IR}$\ relationships carry an 18$\sigma$ and 24$\sigma$ correlation significance, respectively.  Below, we interpret the underlying physics of the correlations. 

To investigate whether the extended redshift range of our sources affects our results, we recalculated the correlations based on a reduced sample with $1.44 \leqslant z \leqslant 2.86$, removing the high-$z$ and low-$z$ extrema from the sample. At $z < 3$, the reduced sample's continuum observations are on the optically thin portion of the Rayleigh-Jeans tail of dust emission (i.e., rest-frame wavelengths longer than $\sim$\,200\,\um), probing cold dust in the ISM. With the redshift-restricted sample, the significance of the \lpeak--$\Sigma_{\rm IR}$ and \lpeak--SFR correlations are reduced somewhat to 16$\sigma$ and 18$\sigma$ significance, respectively, while the correlation between \lpeak\ and \Dms\ and \lpeak\ and sSFR increase to 4$\sigma$ and 8$\sigma$, respectively.  However, the end results are the same: the \lpeak \ correlations with $\Sigma_{\rm IR}$ and SFR are significantly stronger than those with sSFR and \Dms. 

\subsubsection{\lpeak\ vs. $\Sigma_{\rm IR}$}\label{sec:lpeaksigmair}

 We empirically measure the following relationship between \lpeak\ and $\Sigma_{\rm IR}$, or the star-formation surface density, using our sample:
\begin{equation}
 \begin{array}{ll}
    \log(\Sigma_{\rm IR}/[L_\odot\,{\rm kpc}^{-2}]) = & (18.78\pm0.33) \\
    & - (3.80\pm0.16)\log(\lambda_{\rm peak}/\mu\!m)\\
\end{array}
\label{eq:measured_lpeaksigmaIR}
\end{equation}
The overall strength of the correlation is 24$\sigma$.
The slope of this relation is a 2.4$\sigma$ deviation from expectation in Equation~\ref{eq:expected_lpeaksigmaIR}, and the intercept deviates by 4.3$\sigma$.  Comparing to expectation using Wien's Law directly (Equation~\ref{eq:wienslaw_lpeaksigmaIR}), our empirical relation only deviates by 1.3$\sigma$ in the slope and 0.7$\sigma$ in the intercept.

The agreement between Equations~\ref{eq:expected_lpeaksigmaIR}-\ref{eq:wienslaw_lpeaksigmaIR} and Equation~\ref{eq:measured_lpeaksigmaIR} should be somewhat surprising, given that the geometry of galaxies' ISM is quite complex, and that ISM is far from optically thick!  How might we interpret the remarkably close correlation between \lpeak\ and $\Sigma_{\rm IR}$?

Such a tight correlation would be expected if the peak of dust luminosity (traced directly by \lpeak) originates from regions of the ISM that are, indeed, optically thick.  This may not be in direct conflict with more diffuse, cold dust in galaxies' ISM being optically thin, as that dust may not directly dominate the bulk of emission at the {\it peak} of IR emission.

We note that \citet{simpson17} provide an analysis of several (other) high-$z$ DSFGs that have resolved ALMA sizes and measured SED dust temperatures; their Figure 5 shows the relationship between $T_{\rm d}$ and $\Sigma_{\rm IR}$, suggesting tension between predictions from the Stefan-Boltzmann Law and galaxies' dust temperatures, when assumed to be optically thick (they find more agreement when using the best-fit temperatures assuming optically thin SEDs).  The difference in interpretation between our work on that of \citeauthor{simpson17} is the presumed normalization of the Stefan-Boltzmann Law itself and the geometry of dust: their application of the Stefan-Boltzmann Law is drawn from a presumed spherical shell of dust surrounding a single star-forming region (as in Eq~\ref{eq:sphereicalSB}), while our normalization accounts for a more disk-like configuration of a dust screen with several spatially distinct deeply embedded star-forming regions (as in Eq~\ref{eq:sigmairtdust}).  After accounting for the different geometries, we suggest the \citeauthor{simpson17} results, like ours, are quite consistent with a galaxy-scale Stefan-Boltzmann approximation for the emergent IR luminosity.  Given the strength of the correlation in our dataset and the plausible underlying physical interpretation of this relation, we determine that the \lpeak\ vs. $\Sigma_{\rm IR}$ relation is the most fundamental of those we explore in this paper.

We note that the correlation strength of 24$\sigma$ found for \lpeak-$\Sigma_{\rm IR}$\ is unaffected by calculating $\Sigma_{\rm IR}$ with circularized radii or effective radii from Sersic fitting rather than FWHMs from Gaussian fitting. The strength of the correlations are also not affected by removing the two sources (450.03 and 450.14) for which we do not have reliable stellar mass estimates from our analysis.

It is important to note that our measured sizes trace the dust mass as probed on the Raleigh-Jeans tail of blackbody emission in what is expected to be an optically thin regime. This may be a different size than would be probed at the peak SED, which traces SFR density more directly. In other words, one might not expect the mass-weighted size to map directly to the SFR-weighted size. 
 The relative agreement of the two is suggestive of a linear relationship between the effective radius over which the dust mass and star formation exist (i.e., \re \ for SFR is similar to \re \ for \Mdust), which in turn may imply a linear slope for the Kennicutt-Schmidt law of star-formation rate and gas surface density in these galaxies \citep{kennicutt98, delosreyes19}. Further high-resolution follow-up work is needed to explore this in greater detail.

\subsubsection{\lpeak\ vs. $L_{\rm IR}$ or SFR}\label{sec:lpeaklir}

 We measure the following empirical relationship between L$_{\rm IR}$ and \lpeak\ in this sample:
\begin{equation}
    \log(L_{\rm IR}/L_\odot) = (18.2\pm0.3) - (2.81\pm0.16)\log(\lambda_{\rm peak}/\mu\!m)
    \label{eq:lirlpeak}
\end{equation}
This correlation carries an overal significance of 18$\sigma$.  The relationship with SFR can then be derived using the \citet{kennicutt12} scaling.  Figure~\ref{fig:correlations} demonstrates that the trend broadly follows that of the star-formation or IR luminosity surface density, $\Sigma_{\rm IR}$.

We note that this relation is quite a bit steeper than found elsewhere in the literature for much larger samples of galaxies lacking high-resolution dust observations \citep[e.g.][]{lee13a,symeonidis13a,kirkpatrick17a}; for example, \citet{casey18b} modeled the \lir--\lpeak\ relationship by a power law such that $\log_{10}$(\lpeak)$=\log(\lambda_{0}) + \eta(\log(\rm L_{IR})-12)$. Through empirical measurement of large samples of IR luminous galaxies from $0<z<5$, they found $\eta$ = --0.068 $\pm$ 0.001  and $\log(\lambda_{0}/\mu\!m)=2.012$. Though both our findings here and the \citeauthor{casey18a} findings both show the same general trend -- whereby higher luminosity sources are hotter -- here we find the higher luminosity sources are {\it substantially} hotter than those with fainter luminosities.  As touched on earlier in section~\ref{sec:biases}, this is likely due to our selection of sources well-suited for high-resolution dust continuum follow-up (and therefore, relatively high L$_{\rm IR}$ for a wide swath of \lpeak); therefore it is likely that the more extensive samples from \citet{casey18a} and references therein provide a more well-calibrated measurement of the \lpeak-$L_{\rm IR}$ relationship over a much larger dynamic range in \lir.

 Because the  \citet{casey18a} model does not  account for galaxy size as our  analysis in Equation~\ref{eq:measured_lpeaksigmaIR} does,  we can use the two in conjunction to infer an overall size dependency of \lir. Combining our  Equation~\ref{eq:measured_lpeaksigmaIR} with their  \lpeak--\lir\ relationship, we find that the size dependence can be modeled as a power-law function of \lir \ such that  \re\,$\propto$ \lir$^{0.37\pm0.03}$.  Note that this is roughly consistent with the size-luminosity relationship derived in \citet{fujimoto17a}, who found \re\,$\propto$\lir$^{0.28\pm0.07}$.  

\subsubsection{\lpeak\ vs. sSFR or Distance to the Main Sequence}

 Because both specific star-formation rate (sSFR) and the distance of a galaxy from the SFR-\mstar\ relation (the galaxy main sequence) both fundamentally trace physics dependent on {\it both} SFR and stellar mass, we discuss them here together.  While we do find a subtle relationship between \lpeak\ vs. sSFR and \lpeak\ vs. \Dms, they are substantially weaker than direct correlations between \lpeak\ and $\Sigma_{\rm IR}$ or \lir.

In contrast to some literature claims, we do not find  clear evidence to support the theory that galaxies' dust temperatures correlate strongly with a galaxy's position on the galaxy main sequence. We note that the dynamic range of sSFR and \Dms\ in our sample is comparable to similar datasets at matched redshifts ($\sim$2\,dex in sSFR and $\sim$1\,dex in \Dms) (\citet{barger14a}). Specifically, we find the following correlations between \lpeak\ and sSFR or \Dms\ (where we have defined the latter as the orthogonal shortest distance to the \citealt{speagle14a} fits at the source's redshift):
\begin{equation}
\begin{array}{ll}
\log(sSFR/[{\rm Gyr^{-1}}]) = & (3.4\pm0.4) -\\ & (1.43\pm0.20)\log(\lambda_{\rm peak}/\mu\!m)\\
\end{array}
\end{equation}
and 
\begin{equation}
    \log(D_{\rm MS}) = (1.5\pm0.4) - (0.54\pm0.21)\log(\lambda_{\rm peak}/\mu\!m)
\end{equation}
which have overall correlation significance of 7$\sigma$ and 3$\sigma$, respectively.  Note that alternate definitions of \Dms\ (\deltalogssfr\ and/or \Dms\ using the \citealt{whitaker14a} quadratic form of the main sequence) result in similarly low correlation strengths ranging 1--4$\sigma$. 

The fact that both correlations between dust temperature and sSFR or \Dms\ are relatively weak suggests that the inclusion of stellar mass effectively dilutes the underlying physics driving the observed value of \lpeak.  In fact, if we explicitly test for correlation between stellar mass and \lpeak, indeed, we find no significant correlation.

With integrated dust SED temperature lacking strong direct correlation with the main sequence, if the assertion is made that the main sequence informs on the evolutionary stage of a galaxy  then we infer that \tdust\ (or \lpeak) is not  a viable indicator of evolutionary class (starburst or secular disk) and that higher temperature does not necessarily indicate  presence of a starburst \citep{bothwell10a, hodge12a, drew20}.  Rather, our results suggest that the underlying dust geometry within the ISM plays a much more substantial role.  In addition, ISM dust can be heated by AGN whether or not the galaxy hosts an ongoing starburst \citep{kirkpatrick12}. Additional recent cosmological simulation work supports the production of intense luminosities through secularly evolving disk systems either from gas infall, minor mergers, or disk instabilities \citep{dave10a, hodge12a, narayanan15a, hayward18a, tadaki18a}. 

 \citet{magnelli14a} arrived at different results than we have, suggesting significant correlation between \Dms\ and \tdust. Instead of using our definition of \Dms, they evaluated \Dms \ to be \deltalogssfr, where \deltalogssfr = log[SSFR/SSFR$_{\rm MS}$(\mstar,$z$)]. We also investigated the correlation between \deltalogssfr \ and \lpeak \ for our sample and found it to have the same correlation strength as the correlation between \Dms, as defined above, and \lpeak.  Without uncertainties on individual measurements for  \deltalogssfr\ in \citet{magnelli14a}, we are unable to directly compare those results to our analysis on the relative significance of the correlation. We also note that the stacking techniques used in their analysis on {\it Herschel} data may be prone to temperature-dependent biases. For instance, the stacking results of \citet{viero13} result in fundamentally different (and overall colder) SEDs than the stacking results of \citet{bethermin15a} despite use of very similar datasets  (Drew et al. in preparation). Further analysis of SEDs drawn from broad literature sources is beyond the scope of this work.

\subsection{Comparison to local galaxies}

 Note that our findings --- that $\Sigma_{\rm IR}$ correlates most closely with \lpeak\ of a dust SED--- is in agreement with previous work at lower redshifts, for which a relationship between the star formation surface density, $\Sigma_{\rm IR}$, and dust temperature exists \citep[][see also \citealt{diaz-santos10} which shows that nearby galaxies with higher \lir\ have distinctly more compact sizes]{lehnert96a,chanial07,lutz16a}.
While the results of \citet{chanial07} work find a relatively shallow relationship between $\Sigma_{\rm IR}$ and \tdust, our results are more consistent with the earlier (steeper) dependence suggested by \citet{lehnert96a}; the relative discrepancies could be due to measurement of sizes conducted in radio continuum vs. millimeter continuum \citep[of which the literature suggests a discrepancy, e.g.][]{miettinen17a} or differences in the sample selection. 

In particular, we find it useful to analyze the results of \citet{lutz16a} in context, which is based on FIR size measurements from {\it Herschel}/PACS of local galaxies; they derived relationships between $\Sigma_{\rm IR}$, \lir\, and sSFR/sSFR$_{\rm MS}$ and a proxy measure of dust temperature, the FIR color $\log(S_{70}/S_{160})$ which we will call C.  Their Table 2 provides a summary of the derived relationships, allowing us to directly infer the relative correlations between the FIR color $C$ and $\Sigma_{\rm IR}$, \lir\ and \deltalogssfr\ (even though FIR color $C$ is not equivalent to \lpeak).  What the \citeauthor{lutz16a} scaling relations suggest is that $C$ and $\Sigma_{\rm IR}$ are correlated with a measurement significance of 30$\sigma$, $C$ and \lir\ are correlated with a measurement significance of 20$\sigma$, and $C$ and \deltalogssfr\ are correlated with a measurement significance of 6$\sigma$.  While the exact significance of the correlations is likely dependent on sample details, the overall trend is similar to our findings: that $\Sigma_{\rm IR}$ is the most closely linked physical quantity to dust temperature (or \lpeak\ or $C$) and that correlation to the galaxies' main sequence (here probed by \deltalogssfr) is the weakest of the correlations.

\section{Conclusions} \label{sec:conclusion}
We used 0\farcs24 \ resolved 870\,\um \ ALMA dust continuum observations to determine the total integrated flux density and size. We used SED fitting to determine \lpeak \ and \lir. We investigated correlations of dust temperature with four galaxy characteristics: sSFR, parameterized distance to the galaxy main sequence (\Dms), SFR, and star formation surface density measured via $\Sigma_{\rm IR}$. Our results are as follows:
\begin{itemize}
\setlength{\itemsep}{7pt}
\setlength{\parskip}{0pt}
\setlength{\parsep}{0pt} 
\item The \re \ of our seven new DSFGs range from 2.1--12.03 kpc with sizes of the full sample analyzed at 870\,\um \ with $\langle\!$\re$\!\rangle$ = 3.9 $\pm$ 1.6 kpc. The median Sersic index $n = 0.8 \pm 0.4$ implies a non-Gaussian, exponential disk morphology.

\item The correlations of \lpeak \ with sSFR and \Dms \ are relatively weak. The linear models relating $\log$(\lpeak) with $\log$(sSFR) and $\log$(\Dms) had  strengths of only 7$\sigma$ and 3$\sigma$, respectively.

\item The linear models relating $\log$(\lpeak) with $\log$(SFR) and $\log$($\Sigma_{\rm IR}$) had strengths of 18$\sigma$ and 24$\sigma$, respectively, confirming that the \lpeak--\lir \ and \lpeak--$\Sigma_{\rm IR}$\ relationships are statistically stronger than the \lpeak--sSFR and \lpeak--\Dms \ relationships and more likely probe the underlying physical driver of galaxies' luminosity-weighted dust temperatures.

\item Our results are consistent with measured scaling relations between FIR color, $\Sigma_{\rm IR}$, \lir, and $\Delta\log$(sSFR) for galaxies in the local Universe analyzed by \citet{lutz16a}.  In particular, both our work and \citeauthor{lutz16a} find that the \lpeak--$\Sigma_{\rm IR}$ relationships are the strongest.  This motivates our conclusion that SFR surface density is the fundamental characteristic driving a galaxy's luminosity-weighted dust temperature.
\end{itemize}

We conclude that galaxies' dust emission properties can be well described by a Stefan-Boltzmann like law, where the luminosity-weighted dust temperature (as measured via \lpeak) goes roughly as $\Sigma_{\rm IR}\propto$\lpeak$^{-4}$. While there are many physical reasons such a relationship would be expected not to hold (for example, our measurement of size on the Rayleigh Jeans tail of optically thin dust emission, or the fact that overall dust emission is not optically thick throughout the ISM), our finding that it does hold suggests that (a) the bulk of galaxies' IR luminosity can be approximated as optically thick near the SED peak, roughly consistent with a disk or planar geometry, and (b) there is a linear relationship between galaxies' ISM dust-mass weighted sizes and galaxies' SFR-weighted sizes. Furthermore, we infer that dust temperature is not necessarily a reliable indication of where a galaxy sits on the main sequence and should not be used to infer evolutionary class as the quantity is most likely to trace the underlying geometry of galaxies' ISM. 

\acknowledgements{We thank the anonymous reviewer for their input. ADB thanks the John W. Cox Endowment for the Advanced Studies in Astronomy for support. CMC thanks the National Science Foundation for support through grants AST-1714528, AST-1814034, and AST-2009577 and additionally CMC, JAZ, and JSS thank the University of Texas at Austin College of Natural Sciences for support. In addition, CMC acknowledges support from the Research Corporation for Science Advancement from a 2019 Cottrell Scholar Award sponsored by IF/THEN, an initiative of Lyda Hill Philanthropies. SMM thanks the National Science Foundation for support through the Graduate Research Fellowship under Grant No. DGE-1610403. CCC acknowledges support from the Ministry of Science and Technology of Taiwan (MOST 109-2112-M-001-016-MY3). This paper makes use of the following ALMA data: ADS/JAO.ALMA $\#$2015.1.00568.S. ALMA is a partnership of ESO (representing its member states), NSF (USA) and NINS (Japan), together with NRC (Canada), MOST and ASIAA (Taiwan), and KASI (Republic of Korea), in cooperation with the Republic of Chile. The Joint ALMA Observatory is operated by ESO, AUI/NRAO and NAOJ. The National Radio Astronomy Observatory is a facility of the National Science Foundation operated under cooperative agreement by Associated Universities, Inc.}

\bibliography{bibdesk.bib}

\end{document}